\documentclass[a4paper,11pt]{article}
\pdfoutput=1 
\usepackage{jcappub, bm, color} 
\usepackage{amssymb,amsfonts,slashed,amsthm,amsmath,graphicx, soul, empheq}
\usepackage[caption=false]{subfig}
\begin{document}

\newcommand{\vev}[1]{ \left\langle {#1} \right\rangle }
\newcommand{\bra}[1]{ \langle {#1} | }
\newcommand{\ket}[1]{ | {#1} \rangle }
\newcommand{\eV}{ \ {\rm eV} }
\newcommand{\KeV}{ \ {\rm keV} }
\newcommand{\MeV}{\  {\rm MeV} }
\newcommand{\GeV}{\  {\rm GeV} }
\newcommand{\TeV}{\  {\rm TeV} }
\newcommand{\1}{\mbox{1}\hspace{-0.25em}\mbox{l}}
\newcommand{\Red}[1]{{\color{red} {#1}}}

\newcommand{\lmk}{\left(}  
\newcommand{\rmk}{\right)}
\newcommand{\lkk}{\left[}  
\newcommand{\rkk}{\right]}
\newcommand{\lhk}{\left \{ }  
\newcommand{\rhk}{\right \} }
\newcommand{\del}{\partial}  
\newcommand{\la}{\left\langle} 
\newcommand{\ra}{\right\rangle}
\newcommand{\half}{\frac{1}{2}}

\newcommand{\bea}{\begin{array}}
\newcommand{\eea}{\end{array}}
\newcommand{\beq}{\begin{eqnarray}}
\newcommand{\eeq}{\end{eqnarray}}
\newcommand{\eq}[1]{Eq.~(\ref{#1})}

\newcommand{\dd}{\mathrm{d}}
\newcommand{\Mpl}{M_{\rm Pl}}
\newcommand{\mg}{m_{3/2}}
\newcommand{\abs}[1]{\left\vert {#1} \right\vert}
\newcommand{\mphi}{m_{\phi}}
\newcommand{\Hz}{\ {\rm Hz}}
\newcommand{\for}{\quad \text{for }}
\newcommand{\Min}{\text{Min}}
\newcommand{\Max}{\text{Max}}
\newcommand{\Kahler}{K\"{a}hler }
\newcommand{\cphi}{\varphi}
\newcommand{\Tr}{\text{Tr}}
\newcommand{\diag}{{\rm diag}}

\newcommand{\SUf}{SU(3)_{\rm f}}
\newcommand{\Upq}{U(1)_{\rm PQ}}
\newcommand{\Zpq}{Z^{\rm PQ}_3}
\newcommand{\Cpq}{C_{\rm PQ}}
\newcommand{\ubar}{u^c}
\newcommand{\dbar}{d^c}
\newcommand{\ebar}{e^c}
\newcommand{\nubar}{\nu^c}
\newcommand{\Ndw}{N_{\rm DW}}
\newcommand{\Fpq}{F_{\rm PQ}}
\newcommand{\fpq}{v_{\rm PQ}}
\newcommand{\Br}{{\rm Br}}
\newcommand{\Lag}{\mathcal{L}}
\newcommand{\Lqcd}{\Lambda_{\rm QCD}}

\newcommand{\ji}{j_{\rm inf}} 
\newcommand{\jb}{j_{B-L}} 
\newcommand{\M}{M} 
\newcommand{\im}{{\rm Im} }
\newcommand{\re}{{\rm Re} }

\def\lrf#1#2{ \left(\frac{#1}{#2}\right)}
\def\lrfp#1#2#3{ \left(\frac{#1}{#2} \right)^{#3}}
\def\lrp#1#2{\left( #1 \right)^{#2}}
\def\REF#1{Ref.~\cite{#1}}
\def\SEC#1{Sec.~\ref{#1}}
\def\FIG#1{Fig.~\ref{#1}}
\def\EQ#1{Eq.~(\ref{#1})}
\def\EQS#1{Eqs.~(\ref{#1})}
\def\TEV#1{10^{#1}{\rm\,TeV}}
\def\GEV#1{10^{#1}{\rm\,GeV}}
\def\MEV#1{10^{#1}{\rm\,MeV}}
\def\KEV#1{10^{#1}{\rm\,keV}}
\def\blue#1{\textcolor{blue}{#1}}
\def\red#1{\textcolor{blue}{#1}}

\newcommand{\eff}{\Delta N_{\rm eff}}
\newcommand{\neff}{\Delta N_{\rm eff}}
\newcommand{\cc}{\Omega_\Lambda}
\newcommand{\Mpc}{\ {\rm Mpc}}
\newcommand{\Msolar}{M_\odot}

\def\my#1{\textcolor{blue}{#1}}
\def\MY#1{\textcolor{blue}{[{\bf MY:} #1}]}
\def\ft#1{\textcolor{red}{#1}}
\def\FT#1{\textcolor{red}{[{\bf FT:} #1}]}


\begin{flushright}
TU-1116

IPMU20-0132
\end{flushright}

\title{
Trapping Effect for QCD Axion Dark Matter
}

\author{
Shota Nakagawa$^{\spadesuit }$, 
Fuminobu Takahashi$^{\spadesuit, \clubsuit}$, 
Masaki Yamada$^{\diamondsuit,\spadesuit}$
}
\affiliation{$^\spadesuit$ Department of Physics, Tohoku University, 
Sendai, Miyagi 980-8578, Japan
\\*[5pt]
$^\clubsuit$ Kavli IPMU (WPI), UTIAS, 
The University of Tokyo, 
Kashiwa, Chiba 277-8583, Japan \\*[5pt]
$^\diamondsuit$ Frontier Research Institute for Interdisciplinary Sciences, Tohoku University, \\
Sendai, Miyagi 980-8578, Japan

} 

\abstract{
In the early universe, the potential of a scalar field can be significantly modified, and the scalar field may be trapped for a long time in a different location than the current vacuum. The trapping effect can increase or decrease the scalar abundance. For instance, in thermal inflation, a scalar field is trapped at the top of the potential by a thermal effect and dominates the universe to drive inflation for a short period of time. On the other hand, a scalar abundance can be exponentially suppressed in the adiabatic suppression mechanism, where a scalar field moves adiabatically by a time-dependent trapping potential. In this study, we investigate such a trapping effect on the abundance of scalar fields. Specifically, we investigate how the abundance of a scalar field depends on its initial position in the case of a double well potential and identify the physical quantity that controls the abundance. Then, we study the QCD axion abundance for various values of the misalignment angle, where the axon potential receives a large temporal mass due to the Witten effect. We find that the axion abundance is suppressed due to the adiabatic suppression mechanism even when it is trapped near the maximum of the potential, if the  trapping effect is sufficiently large.
}

\emailAdd{shota.nakagawa.r7@dc.tohoku.ac.jp}
\emailAdd{fumi@tohoku.ac.jp}
\emailAdd{m.yamada@tohoku.ac.jp}

\maketitle
\flushbottom

\section{Introduction
\label{sec:introduction}}
In an expanding universe, the vacuum evolves over time. For instance, the electroweak symmetry must have been restored at temperatures above the weak scale when the Higgs field stays at the origin.
In the early universe other scalar fields may also be at different positions than they are today, and
they may be copiously produced when they start to move and oscillate around a minimum of the potential.
Examples of such scalar fields include flatons~\cite{Yamamoto:1985rd,Lyth:1995ka}, moduli~\cite{Coughlan:1983ci,Banks:1993en,deCarlos:1993wie}, string axions, and the QCD axion~\cite{Peccei:1977ur,Peccei:1977hh,Weinberg:1977ma,Wilczek:1977pj,Preskill:1982cy,Abbott:1982af,Dine:1982ah}.

A number of moduli and string axions are known to emerge in connection with the compactification of the extra dimensions in string theory. Some of them are relatively light and have some impacts on cosmological evolution. Such modulus fields are known to cause the so-called cosmological moduli problem~\cite{Coughlan:1983ci,Banks:1993en,deCarlos:1993wie};
a modulus field starts to oscillate about the potential minimum when the mass becomes comparable to the Hubble parameter. Since its initial amplitude is expected to be as large as the Planck scale, it could dominate the universe soon after the onset of oscillations. Its decay could change the abundances of light elements in contradiction with the observation, or produce other long-lived particles such as gravitinos~\cite{Endo:2006zj,Nakamura:2006uc,Dine:2006ii,Endo:2006tf}, supersymmetric particles~\cite{Moroi:1999zb,Endo:2006zj,Nakamura:2006uc,Dine:2006ii}, and axions~\cite{Cicoli:2012aq,Higaki:2012ar,Higaki:2013lra} with a sizable branching fraction. These  particles may spoil the success of the standard cosmology.

A flaton field is introduced to solve the cosmological moduli problem, and it is trapped at the origin by thermal effects until it dominates the universe and drives thermal inflation for a short period. 
Its subsequent decay produces a large amount of entropy, thereby diluting the modulus abundance~\cite{Yamamoto:1985rd,Lyth:1995ka}. It is the thermal mass term caused by interactions with plasma that keeps the flaton at the origin to dominate the universe.

Another solution to the cosmological moduli problem is the adiabatic suppression mechanism~\cite{Linde:1996cx}. The modulus field generically acquires a Hubble-induced mass in the early universe, and if its coefficient is much larger than unity, the modulus field adiabatically tracks the potential minimum as it changes over time, 
and no sizable particle production takes place. In fact, it was shown in Ref.~\cite{Linde:1996cx} that the modulus production in this process can be exponentially suppressed.
It is the large Hubble-induced mass term that traps the modulus field at a different location than the potential minimum in the low energy, and suppresses the modulus abundance.
Note, however,  that some amount of the modulus field is necessarily produced at the end of inflation or during reheating once the dynamics of the inflaton and reheating process were properly taken into account~\cite{Nakayama:2011wqa,Nakayama:2011zy,Nakayama:2008ks,Hagihara:2018uix}, and the total modulus abundance is not  exponentially suppressed.

The QCD axion is a pseudo Nambu-Goldstone boson  in the Peccei-Quinn (PQ) mechanism that solves the strong CP problem~\cite{Peccei:1977ur,Peccei:1977hh,Weinberg:1977ma,Wilczek:1977pj}.
It is a viable candidate for dark matter, which can be produced by the misalignment mechanism~\cite{Abbott:1982af,Dine:1982ah,Preskill:1982cy}.
For the decay constant $f_a$ of order $10^{11 - 12}$\,GeV, the QCD axion can explain all dark matter with the initial misalignment angle $\theta_*$ of order unity. However, for a larger decay constant e.g. $f_a$ of order $10^{16-17}$\,GeV, the initial angle must be less than $\sim 10^{-3}$ to avoid the overproduction. See e.g.~\cite{Steinhardt:1983ia,Lazarides:1990xp,Kawasaki:1995vt,Kawasaki:2004rx,Takahashi:2015waa,Kawasaki:2015lpf,Kawasaki:2017xwt,Kitajima:2014xla,Daido:2015cba,Ho:2018qur,Kitajima:2017peg,Graham:2018jyp,Guth:2018hsa} for various ways to suppress the axion abundance.
In particular, in Refs.~\cite{Takahashi:2015waa,Kawasaki:2015lpf,Kawasaki:2017xwt}, the axion abundance was shown to be exponentially suppressed by a time-dependent PQ breaking potential in a similar mechanism described above. One of such potential arises from the so-called the Witten effect~\cite{Witten:1979ey,Fischler:1983sc} in the presence of monopoles. 

In this paper, we investigate the cosmological effect of trapping scalar fields at a different location than the current potential minimum. As explained above, the trapping effect can increase or decrease the amount of scalar fields, and it is crucial to understand their cosmological role. In the adiabatic suppression mechanism, the scalar potential has been approximated by a simple quadratic potential in the past literature, and it was not clear for what class of potential this mechanism works. In particular, it was not clear for which cases this mechanism becomes ineffective and the scalar abundance gets enhancd as in the case of thermal inflation. In this paper, we investigate the adiabatic suppression mechanism for a more general potential, focusing on the following two cases; (i) the double-well potential and (ii) the QCD axion potential for large values of $\theta_*$.  In both cases, we introduce a large time-dependent mass for the scalar and identify the physical quantity relevant for determining the resultant scalar abundance.

The rest of this paper is organized as follows. In Sec.~\ref{sec:2} we briefly review thermal inflation and the adiabatic suppression mechanism where the scalar abundance can be enhanced or suppressed due to the trapping effect. In Sec.~\ref{sec:3} we study the scalar dynamics in the double-well potential, focusing on how the suppression gets weaker as the initial field position deviates from the minimum and approaches the potential maximum at the origin. In Sec.~\ref{sec:4} we similarly study the QCD axion dynamics in the presence of the Witten effect. The last section is devoted for discussion and conclusions.

\section{The scalar trapping effect
\label{sec:2}}
In this section we briefly review the two cases where the scalar abundance is enhanced or suppressed due to the scalar trapping effect. Specifically we discuss thermal inflation and the adiabatic suppression mechanism. Throughout this paper, we assume spatial homogeneity of the universe as we are interested in the scalar dynamics in the early universe.

\subsection{Thermal inflation}
First we briefly explain the idea of thermal inflation~\cite{Yamamoto:1985rd,Lyth:1995ka}, 
where a flaton field dominates the universe and its decay dilutes the number density of other particles.
Let a real scalar $\phi$ denote the flaton field with the following potential,
\begin{align}
    V(\phi) = -\frac{1}{2}m_\phi^2 \phi^2 + \kappa T^2 \phi^2 + \frac{\phi^{n+4}}{M^2} + {\rm (const.)} \,,
    \label{thermalpot}
\end{align}
where $\kappa$ is a coefficient of the thermal mass, and $M$ is the cut-off scale of the theory. The thermal mass arises from interactions of the flaton $\phi$ with the background plasma. The constant term is chosen so that the potential vanishes at the potential minimum. Here we have assumed that the flaton is trapped by a thermal mass, but it is possible to make use of the Hubble-induced term or something similar for this purpose.

At high temperatures, the universe is assumed to be radiation dominated, and the flaton is trapped at the origin due to the thermal effect. As the universe expands, the radiation energy density decreases, and the flaton potential (including the constant term) comes to dominate the universe, driving thermal inflation. The thermal inflation lasts until the origin is destabilized at $\sqrt{\kappa} T \lesssim m_\phi$, and the typical e-folding number ranges from several to a few tens. 
After thermal inflation, the flaton oscillates about the potential minimum, and decays to the standard model particles, producing a large amount of entropy. Such a late-time entropy production can dilute the moduli abundance. It is the thermal effect that traps the flaton at the origin and make it dominate the universe.

For a comparison to the case for the adiabatic suppression mechanism, let us comment on the case with a Hubble-induced mass term instead of the thermal mass term. Suppose that $\kappa T^2$ is replaced with $C^2 H^2$ in \eq{thermalpot}. Then $\phi$ is trapped at the origin for $C H \gtrsim m_\phi$ and then starts to oscillate around the potential minimum at $C H \sim m_\phi$. In this case, the thermal inflation does not occur but the abundance of $\phi$ is enhanced for a stronger trapping with a larger $C$. The enhancement factor is given by 
\beq
 \frac{\Omega_\phi (C)}{\Omega_\phi (C=1)} \sim C^{3/2} \,, 
 \label{enhance}
\eeq
for $C \gg1$, where $\Omega_\phi$ is the density parameter for $\phi$, and we assume that $\phi$ starts to oscillate in the radiation dominated era. 

\subsection{Adiabatic suppression mechanism}
Now let us explain the adiabatic suppression mechanism~\cite{Linde:1996cx}. 
Suppose that a massive scalar field acquires an effective mass that is proportional to the Hubble parameter such as 
\beq
 V(\phi)=\frac{1}{2}m^2_\phi\phi^2+\frac{1}{2}C^2H^2(\phi-\phi_*)^2\,,
 \label{quadratic}
\eeq
where $C$ ($\ge0$) is a constant that represents the size of the Hubble-induced mass term, $\phi_*$ is its minimum, and $m_\phi$ is the scalar mass at low energy when the Hubble-induced mass is negligible. 
In the radiation dominated era, the Hubble parameter decreases as $H \propto T^2$, and the potential minimum changes with time;
at  high temperatures the minimum is at $\phi\simeq\phi_*$ while at  low temperatures it is at $\phi\simeq0$. 
In the case of $0 \leq C\lesssim 1$, the modulus field is not strongly trapped by the Hubble-induced mass term, and it begins to oscillate around $\phi = 0$ when $H \sim m_\phi$.
In the case of $C\gg1$, on the other hand, the scalar field is trapped so strongly that it begins to track the potential minimum adiabatically when $H\sim m_\phi/C$. 
The resulting abundance of the scalar field is exponentially smaller than the case with $0 \leq C\lesssim 1$. This is the adiabatic suppression mechanism.

The amount of suppression can be evaluated by solving the equation of motion for the homogeneous scalar field: 
\beq
\ddot{\phi}+3H\dot{\phi}+V'(\phi)=0\,.
\label{eom}
\eeq
Here and hereafter, the prime denotes the derivative with respect to the argument of the corresponding function. 
An analytic result for the suppression factor can be obtained from (\ref{eom}) with the initial condition $(\phi, \dot{\phi})=(\phi_*, 0)$ at $H\gg m_\phi$ as~\cite{Linde:1996cx} 
\beq
\frac{\Omega_\phi(C\gg1)}{\Omega_\phi(C=1)}\propto C^{3p+1}\exp\left(-C\pi p \right)\,,
\label{suppression}
\eeq
where $p$ is defined by the time-dependence of the scale factor, $a(t)\propto t^p$, i.e., $H = p/t$. In the radiation dominated era, we have $p=1/2$. Thus, the amplitude of the scalar field is exponentially suppressed for $C \gg 1$. 

The quadratic potential was assumed in the above calculation~\cite{Linde:1996cx, Nakayama:2011wqa}. 
This assumption is considered to be valid for a generic potential if the scalar is initially near the minimum, i.e., if $\phi_* \approx 0$.
However, a singlet scalar field without any special point in its field space is not necessarily near the potential minimum in the early universe, and it could be far away from it.
Thus, it is worth investigating the adiabatic suppression mechanism in a more general case. 
In particular, since the scalar abundance is rather enhanced as \eq{enhance} for the case of the initial position exactly at the top of the potential, it is nontrivial whether the adiabatic suppression mechanism works where the potential is deviated from the quadratic potential, especially in the vicinity of the top of the potential. 
In the subsequent sections, we consider a double-well potential and a periodic potential to investigate the condition for the adiabatic suppression mechanism to work.

\section{A case of double-well potential
\label{sec:3}}
\subsection{Setup
\label{sec:DWabundance}}
\begin{figure}[t!]
\includegraphics[width=10cm]{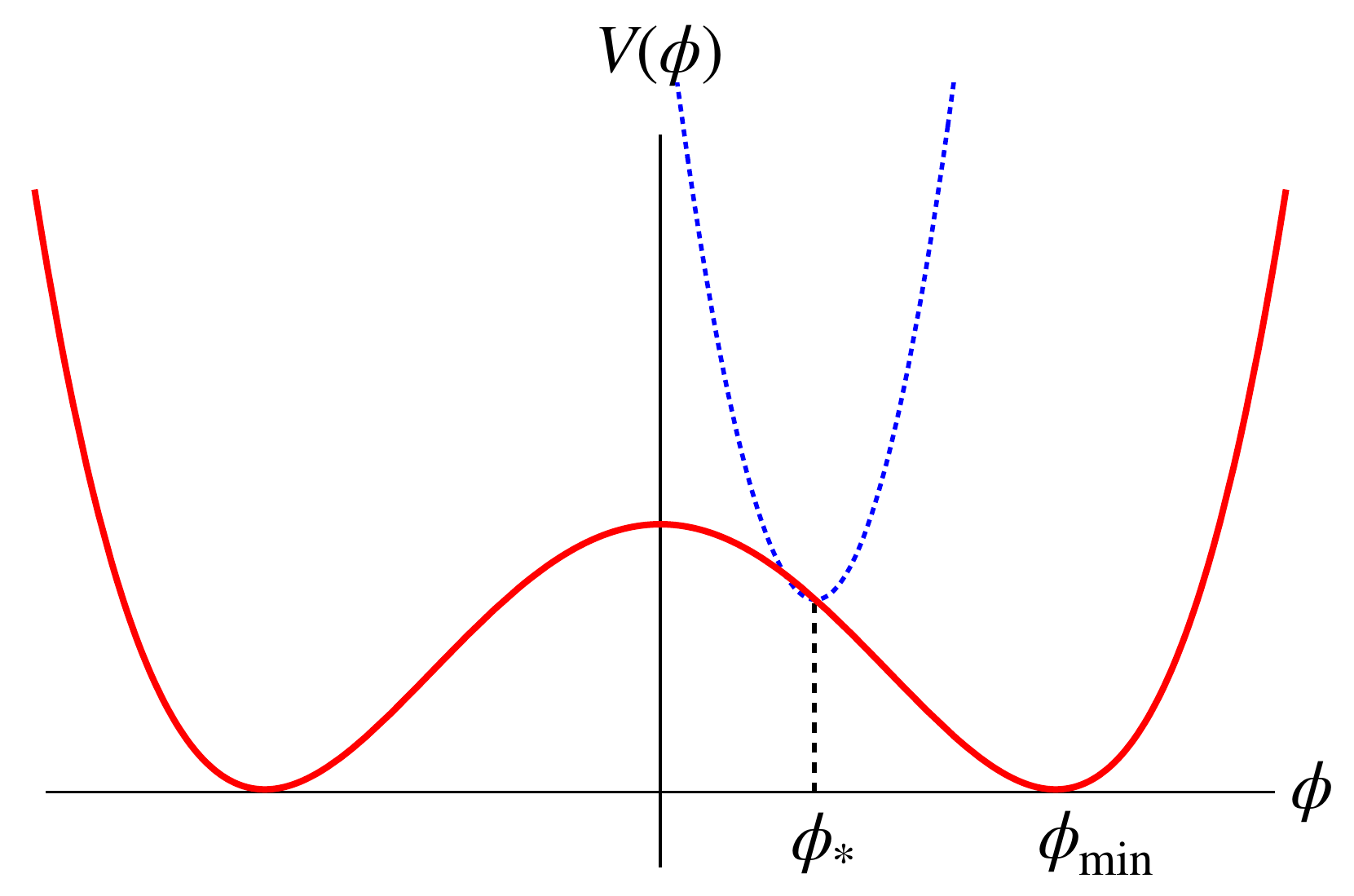}
\centering
\caption{
The schematic picture of the scalar potential. 
The blue dotted line represents the Hubble-induced mass term, which traps the scalar around $\phi_*$ in the early universe.  
The red solid line represents the double-well potential in the late universe.
}
\label{doublewell}
\end{figure}

In order to study if the adiabatic suppression mechanism works near the top of potential, we consider the dynamics of a real scalar field in a double-well potential with a Hubble-induced mass. The potential is given by 
\beq
V(\phi)=\frac{3}{2} \frac{ m_\phi^4}{\lambda}-\frac{1}{2}m^2_\phi\phi^2+\frac{1}{4!}\lambda\phi^4+\frac{1}{2}C^2H^2(\phi-\phi_*)^2\,,
\label{eq:doublewell}
\eeq
where $m_\phi$ is the curvature at the origin, $\lambda$ is a quartic coupling, and $C$ is a constant. 
We assume $m_\phi^2, \lambda >0$ and $C \geq 0$, and focus on the case of $\phi_* > 0$ without loss of generality.
We assume the radiation dominated era when the Hubble parameter is given by
\beq
H^2=g_*(T)\times\frac{\pi^2}{90\Mpl^2}T^4\,. 
\eeq
Here, $\Mpl (\equiv1/\sqrt{8\pi G})$ is the reduced Planck mass and $g_*(T)$ is the effective degrees of freedom of the relativistic particles.
As the universe expands, the Hubble-induced mass becomes smaller, and the minimum of the potential gradually moves from $\phi\simeq\phi_*$ to $\phi \simeq \phi_{\rm min} \equiv \sqrt{6/\lambda}\, m_\phi$. 
See Fig.~\ref{doublewell} for a schematic picture of the potential, where the blue dotted line represents the potential in the early universe, $CH \gg m_\phi$, 
and the red solid line represents the potential at a later time, $CH \ll m_\phi$.
The mass squared of the scalar field at the low-energy minimum is $V''(\phi_{\rm min}) = 2 m_\phi^2$.

To study the dynamics of the scalar field numerically, it is convenient to rewrite the equation of motion in terms of the inverse temperature: 
\beq
 \tau \equiv \frac{T_n}{T}\,, 
\eeq
where $T_n$ is an arbitrary reference temperature. 
We choose $T_n = \sqrt{m_\phi \Mpl}$, which is of order the temperature at the onset of oscillations in the case of $C=0$. 
Note that the Hubble-induced mass becomes comparable to $m_\phi$ when $\tau =\sqrt{C} (g_* \pi^2 / 90)^{1/4} \simeq 1.7 \sqrt{C}$ for $g_* = 80$. 
Since the effective degrees of freedom for energy density $g_*(T)$ and for entropy density $g_{*s}(T)$ depend on the temperature, the time derivative of the temperature is given by
\beq
\frac{dT}{dt}
&=&-\frac{\pi}{\Mpl}\sqrt{\frac{g_*(T)}{10}}\frac{s(T)}{s'(T)}T^2 
\\
&=& - H T \lmk \frac{3\tau F(\tau)}{\sqrt{g_*(\tau)}} \rmk \,.
\label{dTdt}
\eeq
Here, we define 
\beq
F(\tau) \equiv \frac{\sqrt{g_*(\tau)}}{3 \tau} \frac{g_{*s}(\tau)}{g_{*s}(\tau)-\tau g_{*s}'(\tau)/3}\,.
\eeq
Then the equation of motion (\ref{eom}) becomes
\beq
F^2(\tau)\frac{d^2\phi}{d\tau^2}+F(\tau)\left\{\frac{dF(\tau)}{d\tau}+\sqrt{g_*(\tau)}\tau^{-2}\right\}\frac{d\phi}{d\tau}
+\frac{10\Mpl^2}{\pi^2 T_n^4} \del_\phi V(\phi)=0\,. 
\label{generaleom}
\eeq
For the temperature dependence of $g_*$ and $g_{*s}$, the reader may refer to Ref.~\cite{Saikawa:2018rcs}. 
In this section,  we omit the temperature dependence of $g_*$ and $g_{*s}$ for simplicity and set $g_* = g_{*s} = 80$ in our numerical calculation. 
The temperature dependence of $g_*$ and $g_{*s}$ will be taken into account when we consider the dynamics of the QCD axion.

\subsection{Numerical results 
\label{sec:numericalresults}}
For numerical calculations, let us express the field $\phi$ with a dimensionless variable $\theta (\tau)$ defined by
\beq
 \theta (\tau) \equiv \frac{\phi_{\rm min}- \phi(\tau)}{\phi_{\rm min}}\,. 
\eeq
Note that $\phi > 0$ corresponds to $\theta <1$ and the origin $\phi = 0$ corresponds to $\theta = \theta_{\rm max} \equiv 1$. For later convenience we also show the relation,
\beq
\theta-\theta_{\rm max}
= \frac{\phi}{\phi_{\rm min}}\,.
\eeq

We solve the equation of motion numerically under the initial condition $(\theta, \dot{\theta})=(\theta_*, 0)$ at $\tau = \tau_{\rm ini}$, where $\theta_* \equiv \theta \vert_{\phi = \phi_*}$. 
The initial time $\tau_{\rm ini}$ is taken to be $\ll 1$, and we have confirmed that our results do not depend on the choice of $\tau_{\rm ini} \ll 1$. 
Since we are interested in the scalar abundance induced by the shift of the potential minimum, the initial amplitude is set to be zero, i.e., $\theta (\tau_{\rm ini}) = \theta_*$. 
After solving the equation of motion, we calculate the density parameter, which is defined by 
\beq
\Omega_\phi \equiv \frac{\rho_\phi}{\rho_{\rm crit}}\,, 
\eeq
where $\rho_\phi$ is the energy density of the scalar, and $\rho_{\rm crit}$ is the critical density. 
In practice we evaluate the density parameter in terms of the number-to-entropy ratio, $n_\phi / s$, which asymptotes to a constant value at a sufficiently late time.

\begin{figure}[t!]
\includegraphics[width=10cm]{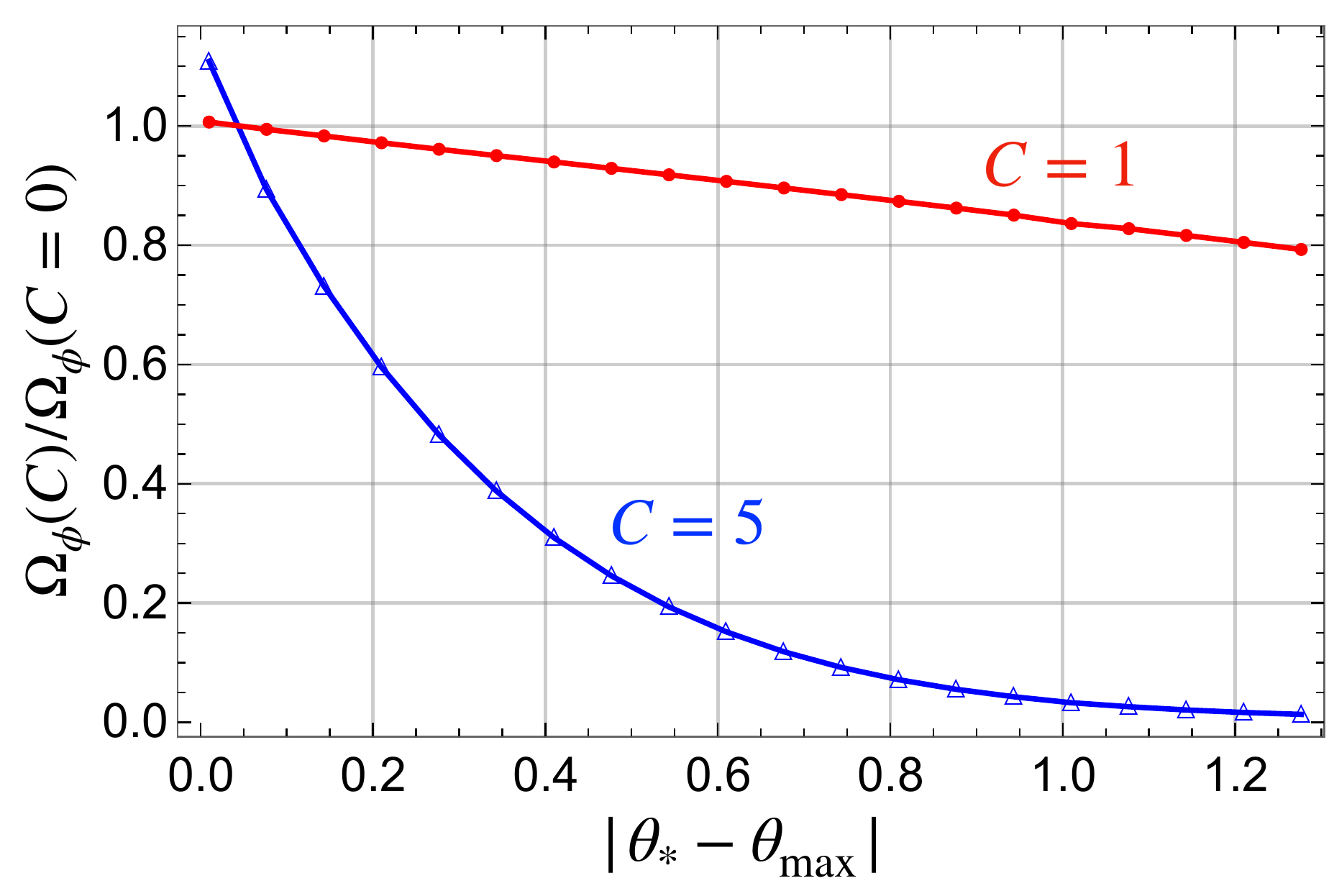}
\centering
\caption{
The density parameter $\Omega_\phi(C)$ normalized by $\Omega_\phi(C=0)$ as a function of the initial position $\abs{\theta_* - \theta_{\rm max}} (=\phi_*/ \phi_{\rm min})$ for $C=1$ (red $\bullet$) and $C=5$ (blue $\triangle$).
Here $\abs{\theta_* - \theta_{\rm max}} = 0$ and $1$ correspond to the origin $\phi_*=0$ and the minimum $\phi_* = \phi_{\rm min}$, respectively.
}
\label{DWratio}
\end{figure}

We have solved Eq.~(\ref{generaleom}) for various values of $\theta_*$, ranging from near the maximum of the double-well potential, $\abs{\theta_* - \theta_{\rm max}} \approx 0$, to the potential minimum $\abs{\theta_* - \theta_{\rm max}} \approx 1$.
In Fig.~\ref{DWratio} we show the density parameter $\Omega_\phi(C)$ normalized by $\Omega_\phi(C=0)$ as a function of the initial position $\abs{\theta_* - \theta_{\rm max}}$ for $C=1$ and $C=5$.
One can see that the abundance is not suppressed in the case of $C=1$, which implies that the abundance is almost the same for $0\leq C \lesssim 1$.
On the other hand, the abundance is significantly suppressed in the case of $C=5$  around $\phi_* = \phi_{\rm min}$ (i.e. $\abs{\theta_* - \theta_{\rm max}} = 1$) where the potential is well approximated by the quadratic potential, while the suppression gets weaker as the initial position approaches the top of the double-well potential $\phi_* = 0$ (i.e. $\abs{\theta_* - \theta_{\rm max}} = 0$). 
This suggests that the adiabatic suppression is weakened near the top of the potential for a given $C$.

\begin{figure}[t!]
\begin{minipage}[t]{8cm}
\includegraphics[width=8cm]{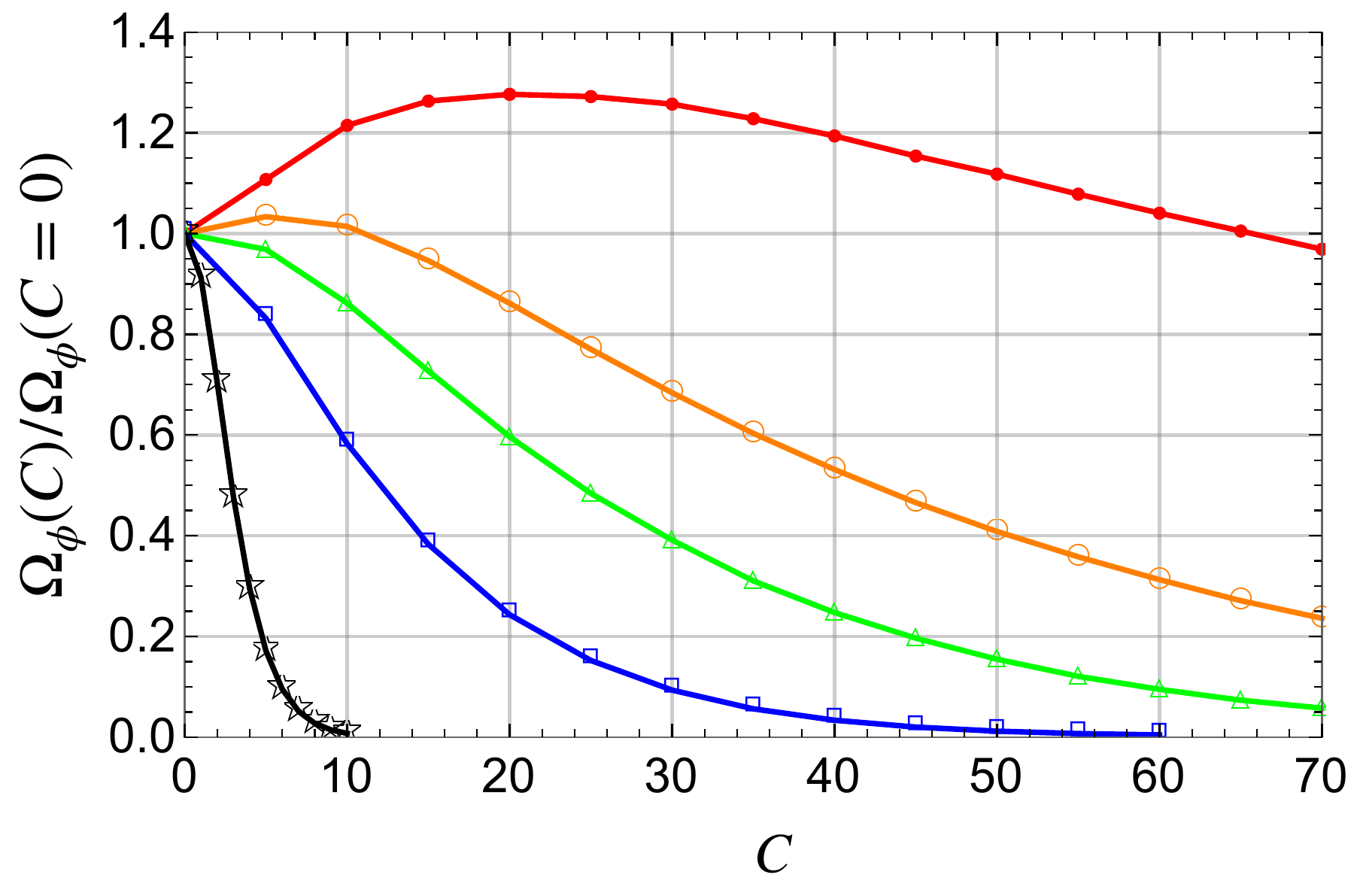}
\centering
\end{minipage}
\begin{minipage}[t]{8cm}
\includegraphics[width=8cm]{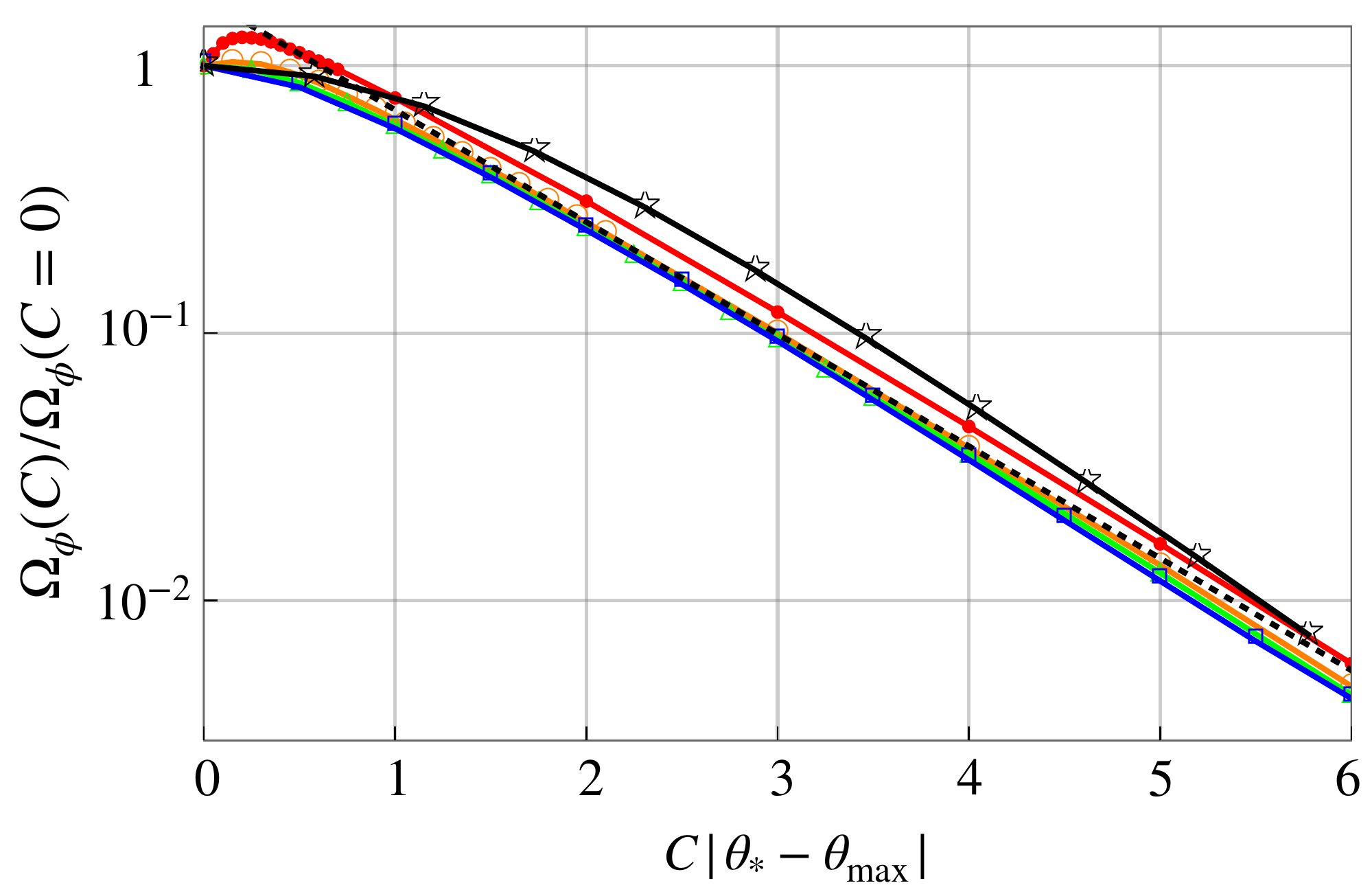}
\centering
\end{minipage}
\caption{
 The density parameter $\Omega_\phi(C)$ normalized by $\Omega_\phi(C=0)$ as a function of $C$ (left) and $C \abs{\theta_* -\theta_{\rm max}}$ (right), with the vertical axis on a linear (left) and logarithmic (right) scale.
 In the left panel, from top to bottom, we vary $\theta_*= 0.01$ (red $\bullet$), $0.03$ (orange $\circ$), $0.05$ (green $\triangle$), $0.1$ (blue $\square$, and $1/\sqrt{3}$(black $\star$).
In the right figure, all the lines exhibit similar suppression as a function of $C \abs{\theta_* -\theta_{\rm max}}$, which can be well fitted by \eq{resultDW} shown by the black dotted line.
}
\label{c-dependence}
\end{figure}

To see how the abundance depends on $C$, we show in  Fig.~\ref{c-dependence} the abundance as a function of $C$ for  $\abs{\theta_* - \theta_{\rm max}}=0.01$, $0.03$, $0.05$, $0.1$, $1/\sqrt{3}$ from top to bottom in the left panel, where the bottom line corresponds the case in which $\theta_*$ is at the inflection point of the double-well potential. 
One can see that, even for the initial condition very close to the potential maximum (as in the case of the top line), the abundance becomes suppressed for a sufficiently large $C$. This implies that it depends on both $C$ and the initial position $\abs{\theta_* - \theta_{\rm max}}$ whether the abundance gets suppressed. 
In the right panel the same result is plotted as a function of $C \abs{\theta_* - \theta_{\rm max}}$, with the vertical axis on a logarithmic scale.
Surprisingly, all the lines exhibit the similar exponential decrease as a function of $C \abs{\theta_* - \theta_{\rm max}}$, and it is well fitted by 
\beq
\frac{\Omega_\phi(C)}{\Omega_\phi(C=0)}\simeq1.8\exp\left(-0.97 C \abs{\theta_*- \theta_{\rm max} } \right)\,,
\label{resultDW}
\eeq
as shown as the black-dotted line in the middle. Thus, the adiabatic suppression works and the abundance gets exponentially suppressed as in the quadratic case (\ref{suppression}). 
However, the exponent in the case of the double-well potential now also depends on the value of the initial field $\phi_*$ (or $\theta_*$), unlike the case of the quadratic potential.
This implies that, although the adiabatic suppression mechanism works for a more general potential than the quadratic one, the anharmonic effect becomes important when the initial position is near 
the top of the potential,
which limits the amount of suppression for a given $C$. We will show that the exponent coincides with a parameter for the violation of adiabaticity in the next subsection.

\begin{figure}[t!]
\begin{minipage}[t]{8cm}
\includegraphics[width=8cm]{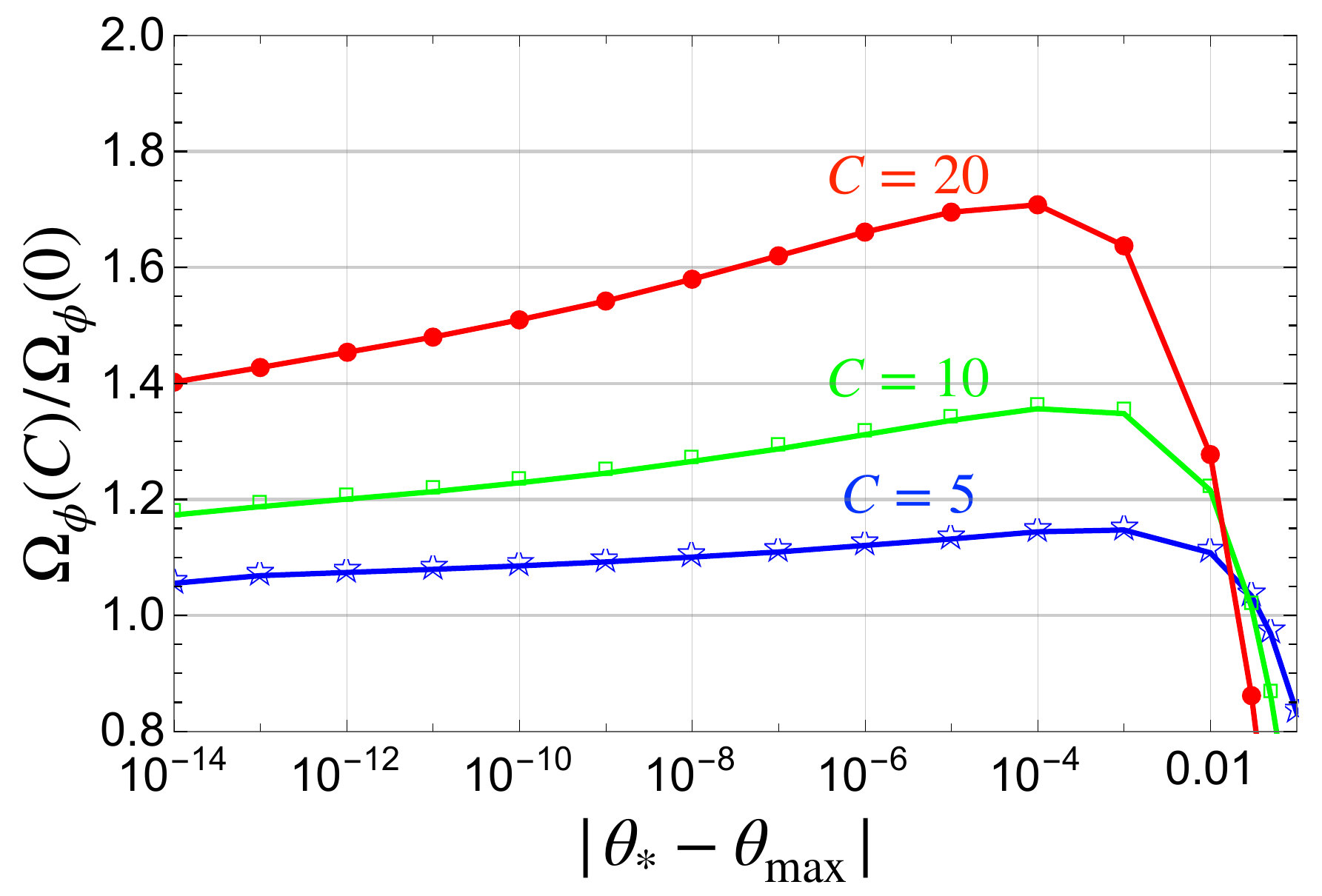}
\centering
\end{minipage}
\begin{minipage}[t]{8cm}
\includegraphics[width=8cm]{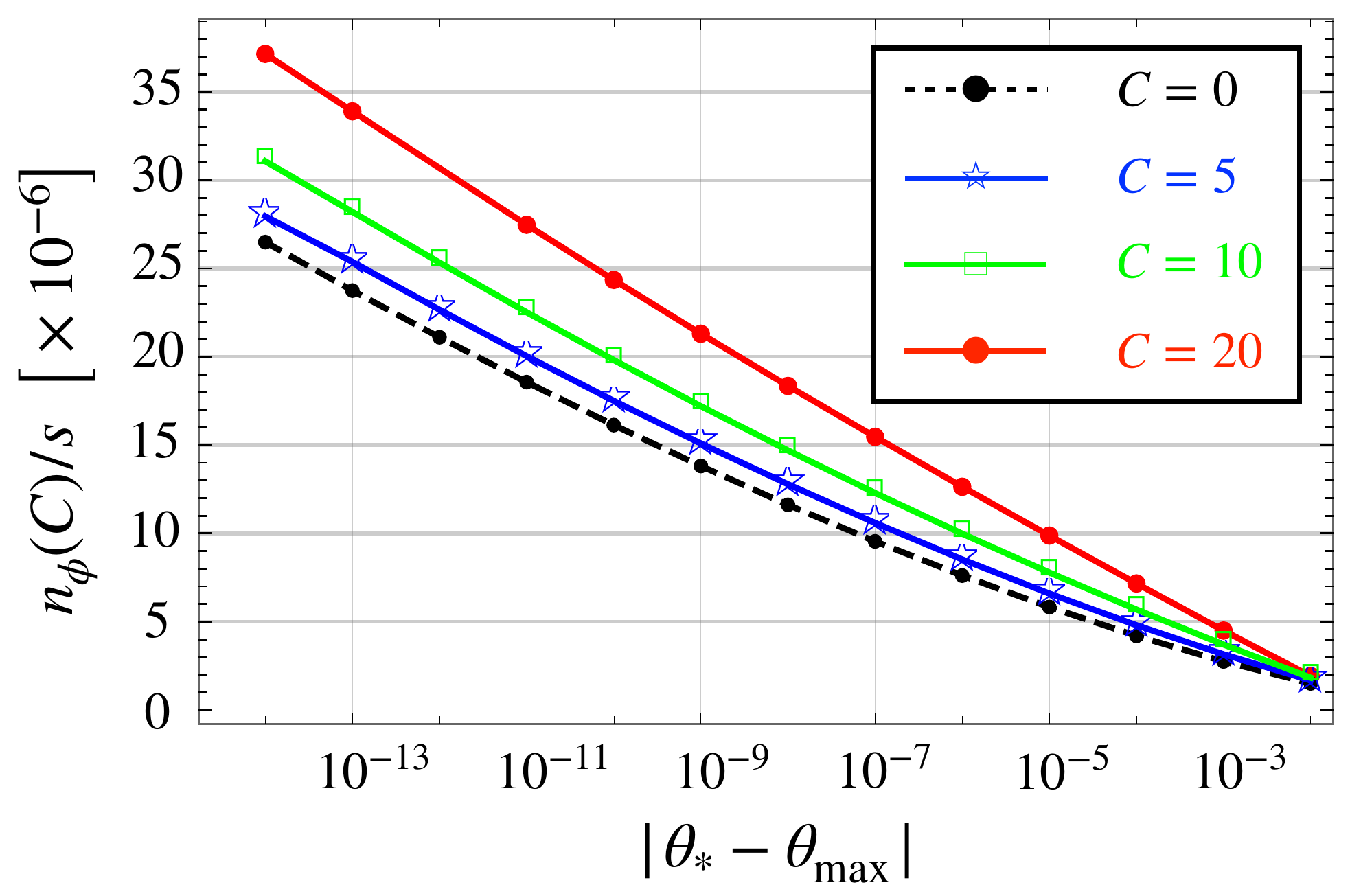}
\centering
\end{minipage}
\caption{
 The density parameter $\Omega_\phi(C)$ normalized by $\Omega_\phi(C=0)$ (left) and the number density $n_\phi (C) / s$ (right) as a function of $\abs{\theta_* -\theta_{\rm max}}$ on the logarithmic scale.
 In the left panel, from top to bottom, we vary $C= 20$ (red $\bullet$), $10$ (green $\square$), and $5$ (blue $\star$). 
 In the right panel, we also show the case with $C = 0$ as a black dashed line with dots. 
}
\label{theta-dependence}
\end{figure}

We also show the abundance as a function of $\abs{\theta_* - \theta_{\rm max}}$ for $C = 5, 10$, and $20$ in Fig.~\ref{theta-dependence}, focusing on a very small $\abs{\theta_* - \theta_{\rm max}}$. 
One can see that, for the initial condition very close to the potential maximum with a large (but not too large) $C$, the abundance is larger than the one without the trapping potential. However, the abundance only logarithmically depends on $C$ as well as $\abs{\theta_* - \theta_{\rm max}}^{-1}$. 
The latter dependence is actually expected because a scalar field cannot stay near the top of the double-well potential for a long time even if the initial field value is very close to the top. 
In the left figure, we find that the enhancement factor decreases as $\abs{\theta_* - \theta_{\rm max}}$ decreases with a fixed $C$ for a very small $\abs{\theta_* - \theta_{\rm max}}$. This is because, for such a small $\abs{\theta_* - \theta_{\rm max}}$, the scalar field stays near the top of the potential until the trapping potential disappears and its effect becomes irrelevant. 
This implies that the effect of the trapping potential is limited and the resulting enhancement factor is not sizable even for a large $C$.

Before closing this subsection, let us discuss the typical time evolution of the scalar for different initial conditions and $C$.
One can see from the left panel of Fig.~\ref{c-dependence} that for $\theta_* = 0.01$ and $0.03$ (top two lines),  the scalar abundance initially increases and then starts to decrease as $C$ increases.
On the other hand, for a fixed $C$, the abundance decreases as $\theta_*$ increases (from top to bottom).
These features can be understood by noting that the scalar abundance is induced by the shift of the temporal minimum. In other words, the scalar abundance crucially depends on whether the scalar field can track the temporal minimum.
The location of the temporal minimum for $\abs{\theta_* - \theta_{\rm max}} \ll 1$ can be obtained by completing the square of the potential as
\beq
V(\phi)=\frac{1}{2} (C^2H^2(t)-m_\phi^2)\left(\phi-\phi_{\rm min}^{\rm temp}(t) \right)^2+\cdot\cdot\cdot \,,
\label{completion}
\eeq
where we have defined the temporal minimum of the potential at a given time, $\phi_{\rm min}^{\rm temp}(t)$, by
\beq
 \phi_{\rm min}^{\rm temp}(t) \equiv \frac{C^2H^2(t)}{C^2H^2(t)-m_\phi^2}\phi_*\,. 
\eeq
At $H\gg m_\phi/C$, the temporal minimum is close to $\phi_*$, and when $H \sim m_\phi/C$ , i.e.  $\tau \sim 1.7 \sqrt{C}$, it starts to move. If the scalar can follow the motion of the temporal minimum, the induced oscillation amplitude remains small, and the final scalar abundance is suppressed. 
Otherwise, the scalar field will be left behind by the motion of the temporal minimum and will catch up with it later. Therefore, the scalar field will start oscillating with a large initial amplitude, and the final abundance is not suppressed. 

\begin{figure}[t!]
\begin{minipage}[t]{8cm}
\includegraphics[width=8cm]{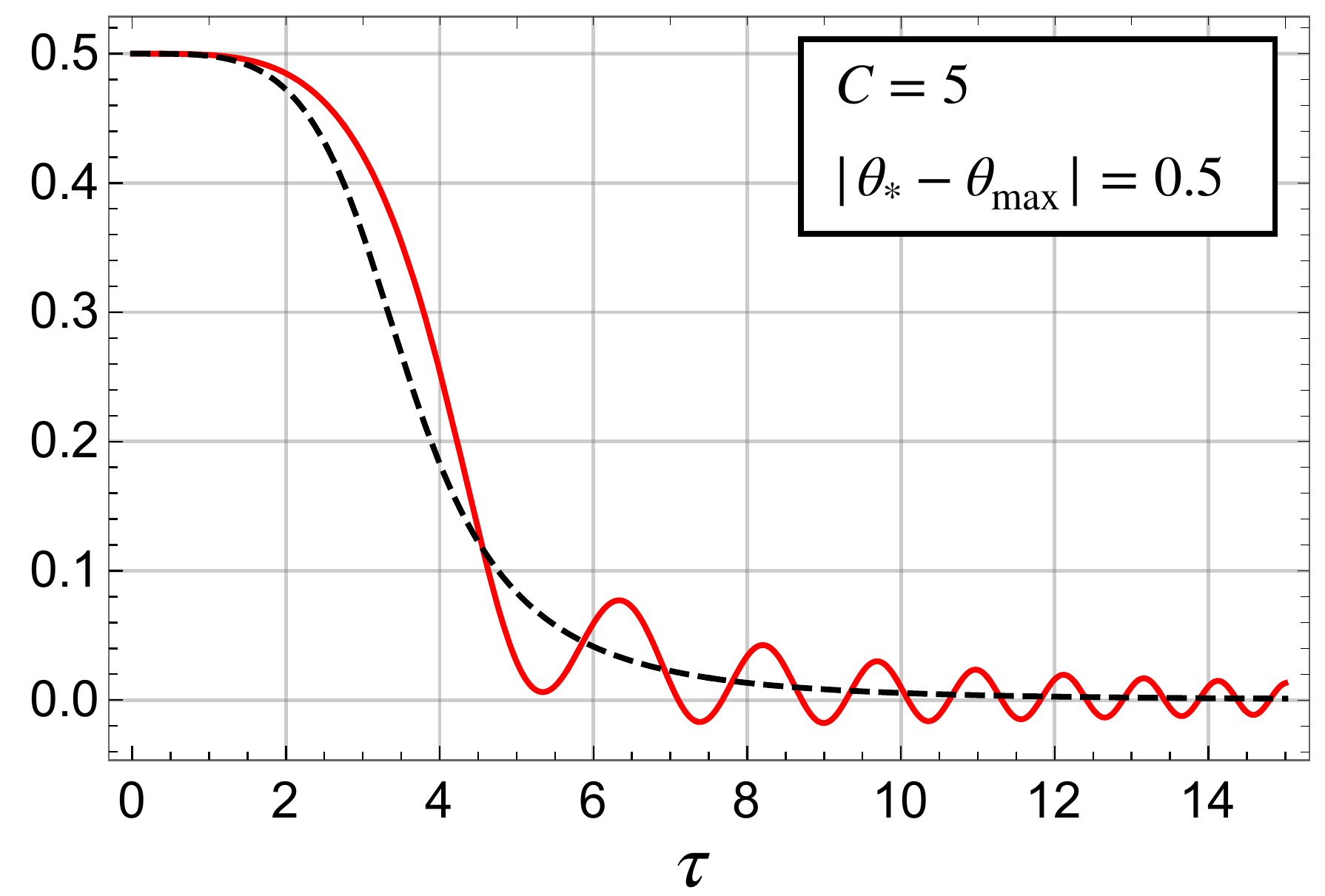}
\centering
\end{minipage}
\begin{minipage}[t]{8cm}
\includegraphics[width=8cm]{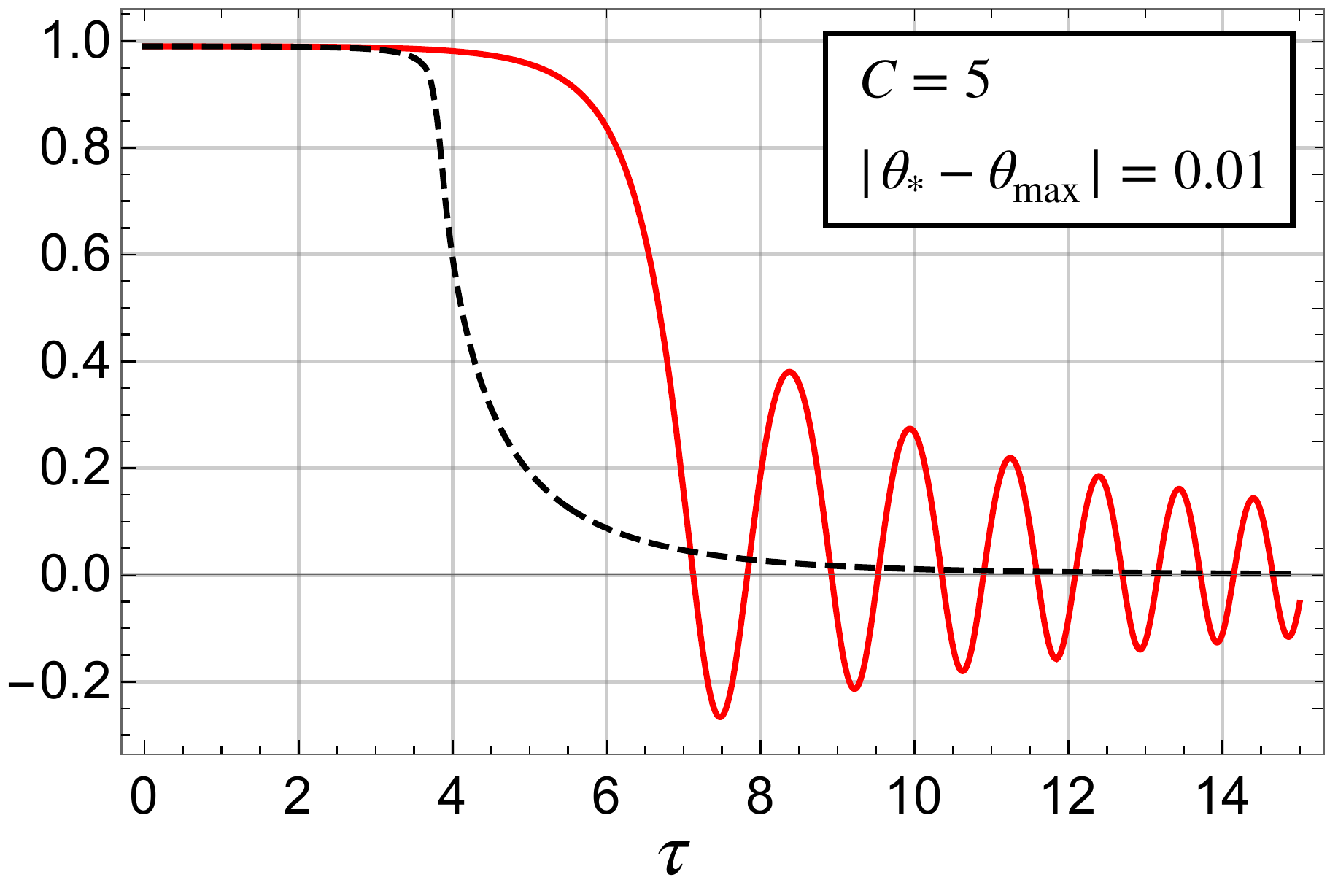}
\centering
\end{minipage}
\caption{
Time evolution of the scalar field (red solid) and the temporal minimum (black dashed) for $C=5$ with $|\theta_*-\theta_{\rm{max}}|=0.5$ (left) and $0.01$ (right). In the right panel, the temporal minimum changes more rapidly, which results in a larger oscillation amplitude.
}
\label{vev-field5}
\end{figure}

\begin{figure}[t!]
\includegraphics[width=10cm]{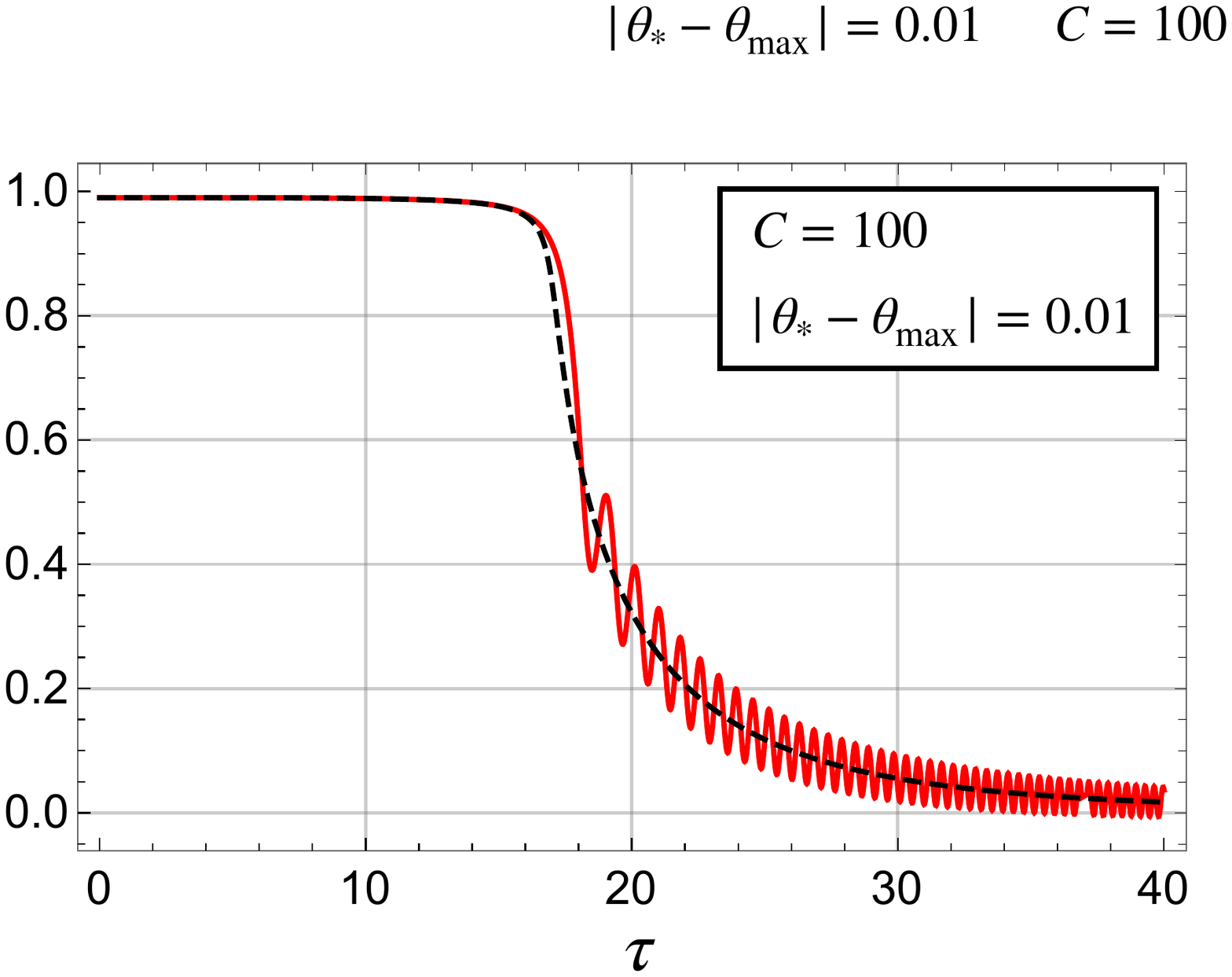}
\centering
\caption{
Same as Fig.~\ref{vev-field5} but with $C=100$ and $|\theta_*-\theta_{\rm{max}}|=0.01$. The onset of oscillation is delayed due to large $C$, but the scalar can track the motion of the minimum compared to the right panel of Fig.~\ref{vev-field5}.
}
\label{vev-field50}
\end{figure}

In  Fig.~\ref{vev-field5} we show the time evolution of the scalar field by the red solid line and the temporal minimum by the black dashed line. Here we set $C=5$ with $\abs{\theta_* - \theta_{\rm max}} = 0.5$ (left) and $0.01$ (right), respectively. In both cases the temporal minimum starts to move when $\tau \sim 1.7 \sqrt{C} \sim 4$.
One can see that the temporal minimum changes more rapidly for smaller  $\abs{\theta_* - \theta_{\rm max}}$. This makes it difficult for the scalar to track the temporal minimum as the initial position comes close to the origin. This is the reason why the scalar abundance increases as $\abs{\theta_* - \theta_{\rm max}}$ decreases for a given $C$. However, such enhancement can be mitigated by increasing $C$. In fact, increasing $C$ has two different effects on the scalar abundance.
In Fig.~\ref{vev-field50} we show the case with $C = 100$ and $\abs{\theta_* - \theta_{\rm max}} =0.01$. 
The minimum starts to move around  $\tau \sim 1.7 \sqrt{C} \sim 17$. Thus, compared to the case of $C=5$ in Fig.~\ref{vev-field5}, the onset of oscillation is delayed. 
One can also see that the scalar can follow the motion of the temporal minimum more closely. This implies that, even though the onset of oscillation is delayed, the abundance may be suppressed for a sufficiently large $C$. 
We will see in the next subsection that this is indeed the case, and it is $C \abs{\theta_* - \theta_{\rm max}}$ that describes the violation of adiabaticity and determines the final scalar abundance.

\subsection{Analytical evaluation of violation of adiabaticity
\label{sec:adiabatic}}
Here we shall explain how the exponent of \eq{resultDW} is derived analytically.
To this end, we introduce a parameter that describes the violation of adiabaticity, $\epsilon(t)$, defined by\footnote{
Alternatively one can use the potential minimum as a parameter for the violation of adiabaticity, $\abs{(d \ln \phi_{\rm min}^{\rm temp}/dt) / m_{\rm{eff}} }$, which is actually of the same order with $\epsilon$.
} 
\beq
 \epsilon(t) \equiv \abs{\frac{1}{m_{\rm{eff}}^2(\phi)} \frac{d {m}_{\rm{eff}}(\phi)}{dt}}\,, 
 \label{adiabaticity}
\eeq
where the effective mass is defined by $m^2_{\rm{eff}}\equiv |V''(\phi)|$. 
If $\epsilon \ll 1$, the effective mass changes adiabatically, which implies that the system also evolves adiabatically, i.e., the scalar field tracks the temporal minimum, as long as the effective mass is larger than the Hubble parameter.  In the following we evaluate $\epsilon$ at $\tau\sim 1.7\sqrt{C}$, since it is expected to reach its maximum value  when the potential minimum starts to move.

To calculate $\epsilon(t)$, we focus on the scalar dynamics with the initial condition near the top of the double-well potential, $\phi_* \ll \phi_{\rm min}$, before the onset of oscillations, $H > m_\phi/C$. 
Then, the potential can be well approximated by
\beq
V(\phi) = \frac12 \lmk C^2 H^2(t) - m_\phi^2 \rmk \lmk \phi - \phi_{\rm min}^{\rm temp} (t)\rmk^2 
+ \frac{1}{24} \lambda \phi^4
+ ({\rm const.})\,. 
\eeq
The effective mass squared at $\phi$ is given by 
\beq
 m_{\rm{eff}}^2(\phi) = C^2 H^2(t) - m_\phi^2 + \frac{\lambda}{2} \phi^2\,,
\eeq
where we have removed the absolute value bars.
As the time $t$ approaches $C / 2 m_\phi$, the first two terms in $m_{\rm{eff}}^2(\phi)$ get almost canceled, and the third term becomes relevant. 
If we take $\phi = \phi_{\rm min}^{\rm temp}(t)$, the threshold time $t_{\rm th}$ after which the third term becomes relevant is given by 
\beq
 C^2 H^2_{\rm th} - m_\phi^2 &=& \lmk \frac{\lambda}{2} C^4 H_{\rm th}^4 \phi_*^2 \rmk^{1/3}
 \\
 &\simeq& \lmk \frac{\lambda}{2} m_\phi^4 \phi_*^2 \rmk^{1/3}\,,
 \label{eq1}
\eeq
where $H_{\rm th} \equiv 1/2t_{\rm th} $ ($\simeq m_\phi / C$).

Let us assume that the scalar tracks the temporal minimum adiabatically until $t = t_{\rm th}$. Then $\epsilon(t)$ is calculated as 
\beq
\epsilon(t)
 &=& \abs{ \frac{1}{\lmk m_{\rm{eff}}^2(\phi_{\rm min}^{\rm temp}(t)) \rmk^{3/2}} \lmk \frac{C^2H^2(t)}{t} - \frac{\lambda \phi_*^2}{t} 
 \frac{m_\phi^2 C^4 H^4(t)}{\lmk C^2 H^2(t) - m_\phi^2 \rmk^3} \rmk}\,, 
\eeq
for $t \lesssim t_{\rm th}$. One can show that $\epsilon(t)$ becomes the largest at $t \sim t_{\rm th}$ and 
\beq
 \epsilon(t_{\rm th})
 &\simeq& \frac{m_\phi}{2 \sqrt{\lambda} C \phi_*}
 \\
 &=& \frac{1}{2 \sqrt{6} C \abs{\theta_* - \theta_{\rm max}}}\,, 
\eeq
where we have used \eq{eq1} and $\phi_* /\phi_{\rm min} = \abs{\theta_* - \theta_{\rm max}} $. 
For $t < t_{\rm th}$, therefore, $\epsilon(t)$ is always smaller than unity  if $C \abs{\theta_* - \theta_{\rm max}} \gtrsim {\cal O}(1)$.
For $t > t_{\rm th}$, the movement of the temporal minimum becomes progressively slower, and the temporal minimum asymptotically approaches the minimum at low energy. Thus, for the initial condition near the origin, we expect that $\epsilon(t)$ takes  the largest value when $t \sim t_{\rm th}$ during the whole evolution, and the adiabatic suppression mechanism works if $\epsilon(t_{\rm th}) \ll 1$.

It is known that the change of the adiabatic invariant is exponentially suppressed as a function of the violation of the adaibaticity~\cite{Landau-mechanics}. In our case, the resulting scalar abundance is exponentially suppressed by a factor of $e^{-{\cal O}(1) / \epsilon(t_{\rm th})} = e^{-\zeta C \abs{\theta_* - \theta_{\rm max}}}$ with $\zeta$ being a numerical constant of order unity. 
This is consistent with \eq{resultDW}, which can be well fitted with $\zeta \simeq 0.97$. 

\begin{figure}[t!]
\begin{minipage}[t]{8cm}
\includegraphics[width=8cm]{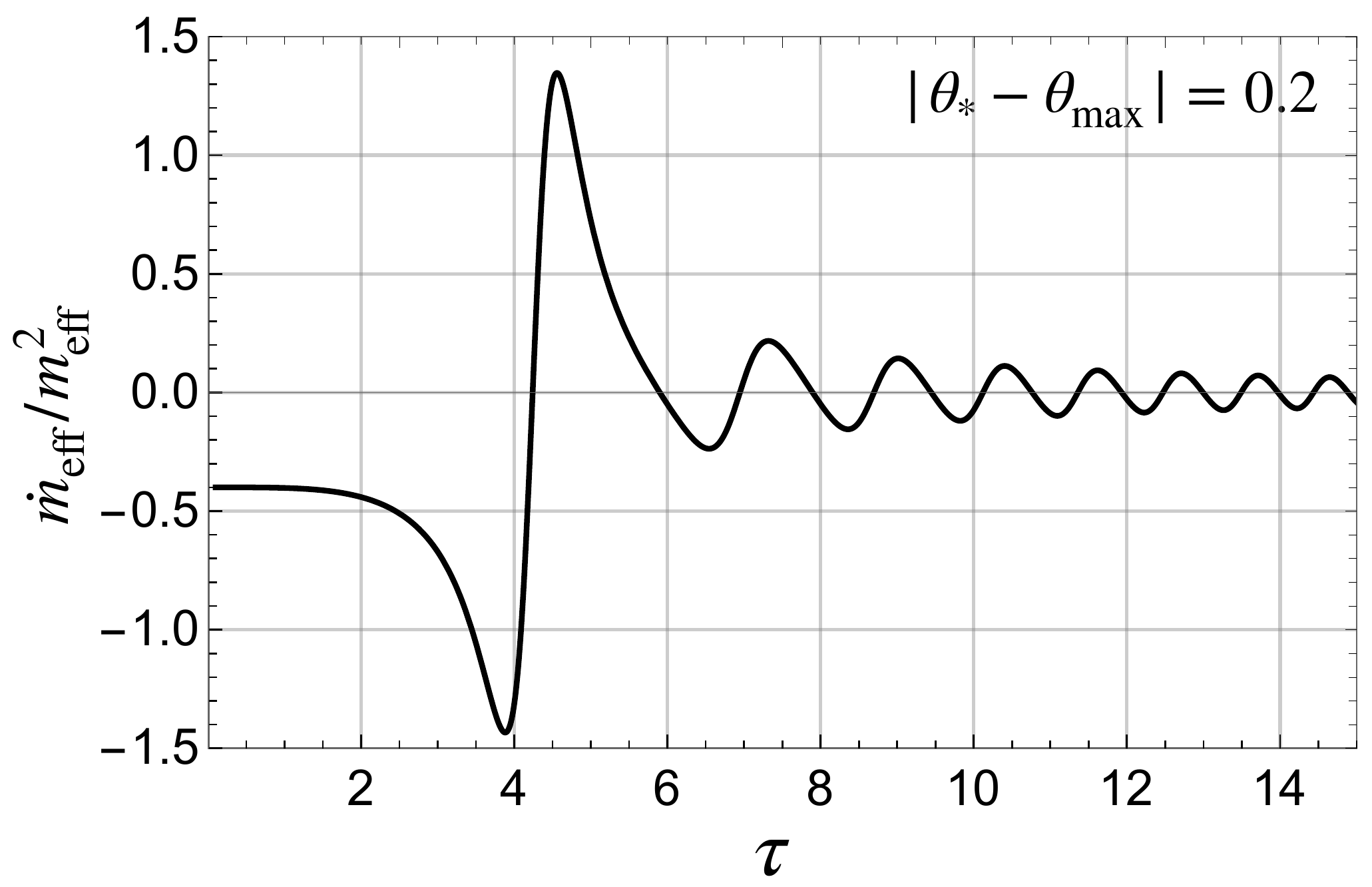}
\centering
\end{minipage}
\begin{minipage}[t]{8cm}
\includegraphics[width=8cm]{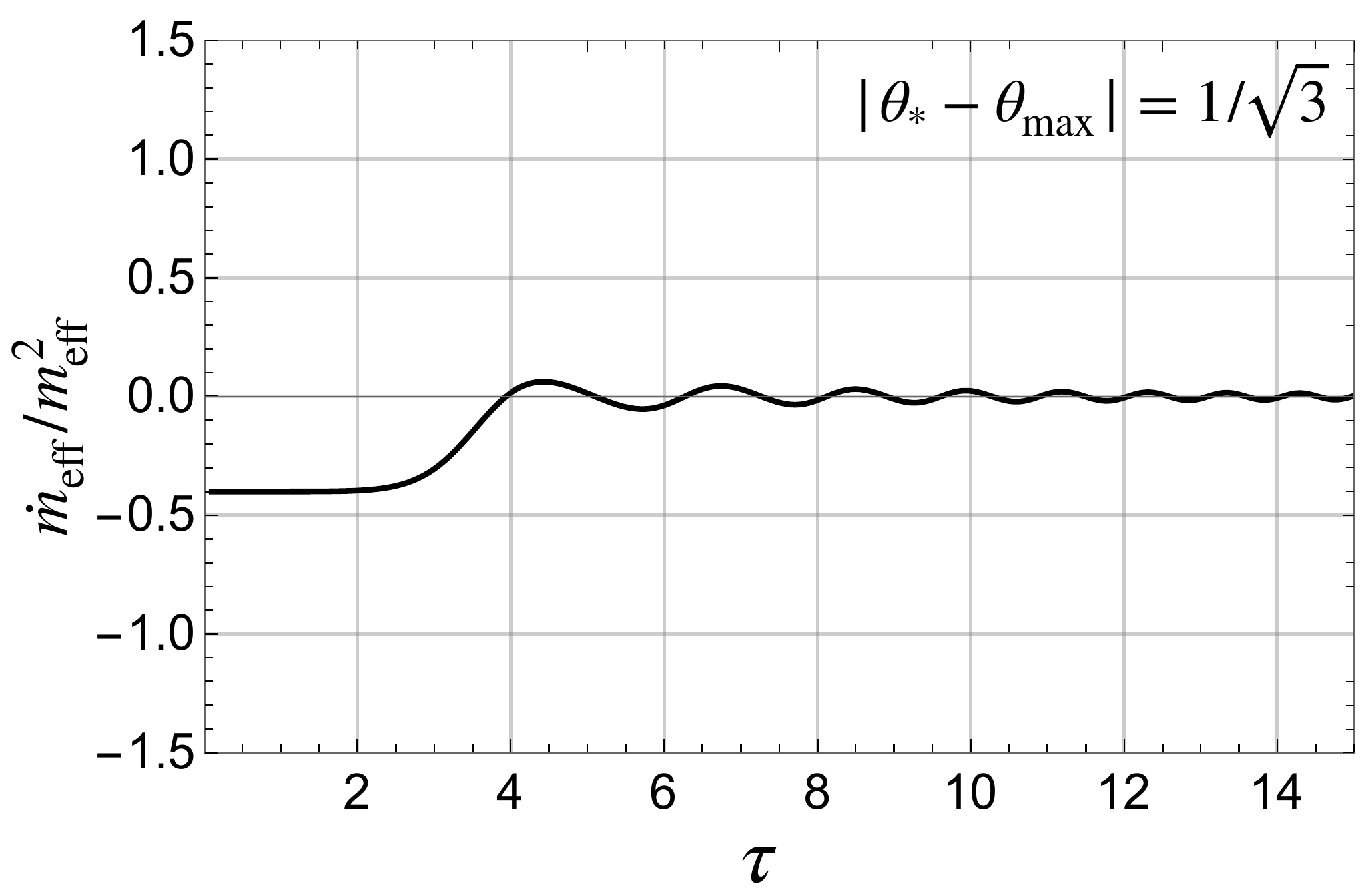}
\centering
\end{minipage}
\caption{
Time evolution of the violation of the adiabaticity, $\dot{m}_{\rm{eff}}(\phi)/m_{\rm{eff}}^2$, for the case of $C=5$ and $|\theta_* - \theta_{\rm max}| = 0.2$(left) and $1/\sqrt{3}$(right). 
The violation of adiabaticity is stronger for smaller $\abs{\theta_* - \theta_{\rm max}}$.
}
\label{fig:adiabaticity}
\end{figure}

We show in Fig.~\ref{fig:adiabaticity}   the time-evolution of $(d{m}_{\rm{eff}}(\phi) / dt) /m_{\rm{eff}}^2(\phi)$ for the case of $C=5$ with $|\theta_* - \theta_{\rm max}| = 0.2$ (left) and  $1/\sqrt{3}$ (right). In both cases the onset of oscillation is around $\tau \simeq 4$ as before. In the left panel, one can see that $\epsilon(t)$ (the absolute value of the vertical axis) takes the largest value around $\tau \simeq 4$  as expected. On the other hand, the right panel corresponds to the case where the potential can be approximated by the quadratic potential. 
In this case the violation of adiabaticity always remains weak, and the abundance is exponentially suppressed (see Fig.~\ref{c-dependence}).

\section{A case of QCD axion with the Witten effect
\label{sec:4}}
The PQ symmetry must be of high quality for the PQ mechanism to solve the strong CP problem. However, it is possible that the axion acquires a temporal mass in the ealy universe, which is much larger than the present axion mass.
Such a temporal potential can reduce the axion abundance and/or isocurvature perturbation and solve the domain wall problem, and it may arise from e.g. stronger QCD~\cite{Dvali:1995ce}, a non-minimal coupling to gravity~\cite{Folkerts:2013tua,Takahashi:2015waa}, and the Witten effect~\cite{Witten:1979ey,Fischler:1983sc}.
The adiabatic suppression mechanism was applied to the axion dynamics in the presence of the Witten effect in Refs.~\cite{Kawasaki:2015lpf,Nomura:2015xil,Kawasaki:2017xwt,Sato:2018nqy}, where  the  potential was approximated by a quadratic term. 
However, the axion potential is periodic, and, as we have seen in Sec.\ref{sec:3}, it is non-trivial whether the adiabatic suppression mechanism works where the potential is concave upward e.g. around the top of the potential.
In this section, we investigate the adiabatic suppression mechanism for the QCD axion with an effective potential that comes from the Witten effect in the presence of hidden monopoles.

\subsection{Axion potential
\label{sec:axion potential}}

First we briefly summarize the axion potential we use in this paper. There are two contributions; one is from non-perturbative QCD effects, and the other from the Witten effect. We explain these two contributions  in turn.

\subsubsection{Non-perturbative effects of QCD
\label{sec:QCDpotential}}
The QCD axion acquires an effective potential from non-perturbative effects of QCD \cite{tHooft:1976rip, tHooft:1976snw}.
At temperatures below the QCD phase transition, i.e., at $T\ll T_{\rm{QCD}}=\mathcal{O}(100)\MeV$, we can evaluate the potential using the chiral perturbation theory.
Considering the two flavor effective field theory \cite{diCortona:2015ldu}, the potential at zero temperature is given by 
\beq
V(a)=-m_\pi^2f_\pi^2\sqrt{1-\frac{4m_um_d}{(m_u+m_d)^2}\sin^2\left(\frac{a}{2f_a}\right)}\,,
\label{chiralpot}
\eeq
where $f_a$ is the axion decay constant, $f_\pi=92.21\MeV$ is the pion decay constant, $m_\pi\simeq135\MeV$ is the pion mass, and $m_u$ and $m_d$ denote the mass of the up- and down-quark. The mass ratio of the up and down quarks is given by $z \equiv m_u/m_d \simeq0.48$ from the average of the lattice results~\cite{deDivitiis:2013xla, Basak:2015lla, Horsley:2015eaa}.
The axion mass is given by 
\beq
m_{a, 0}
&=&\frac{\sqrt{z}}{1+z}\frac{f_\pi m_\pi}{f_a}
\\
&\simeq&5.7\mu\hspace{-1mm}\eV\left(\frac{10^{12}\GeV}{f_a}\right)\,,
\label{NLOmass}
\eeq
where in the second line we quote the result of the next-to-next-to-leading order in chiral perturbation theory~\cite{Gorghetto:2018ocs}.

\begin{figure}[t!]
\includegraphics[width=10cm]{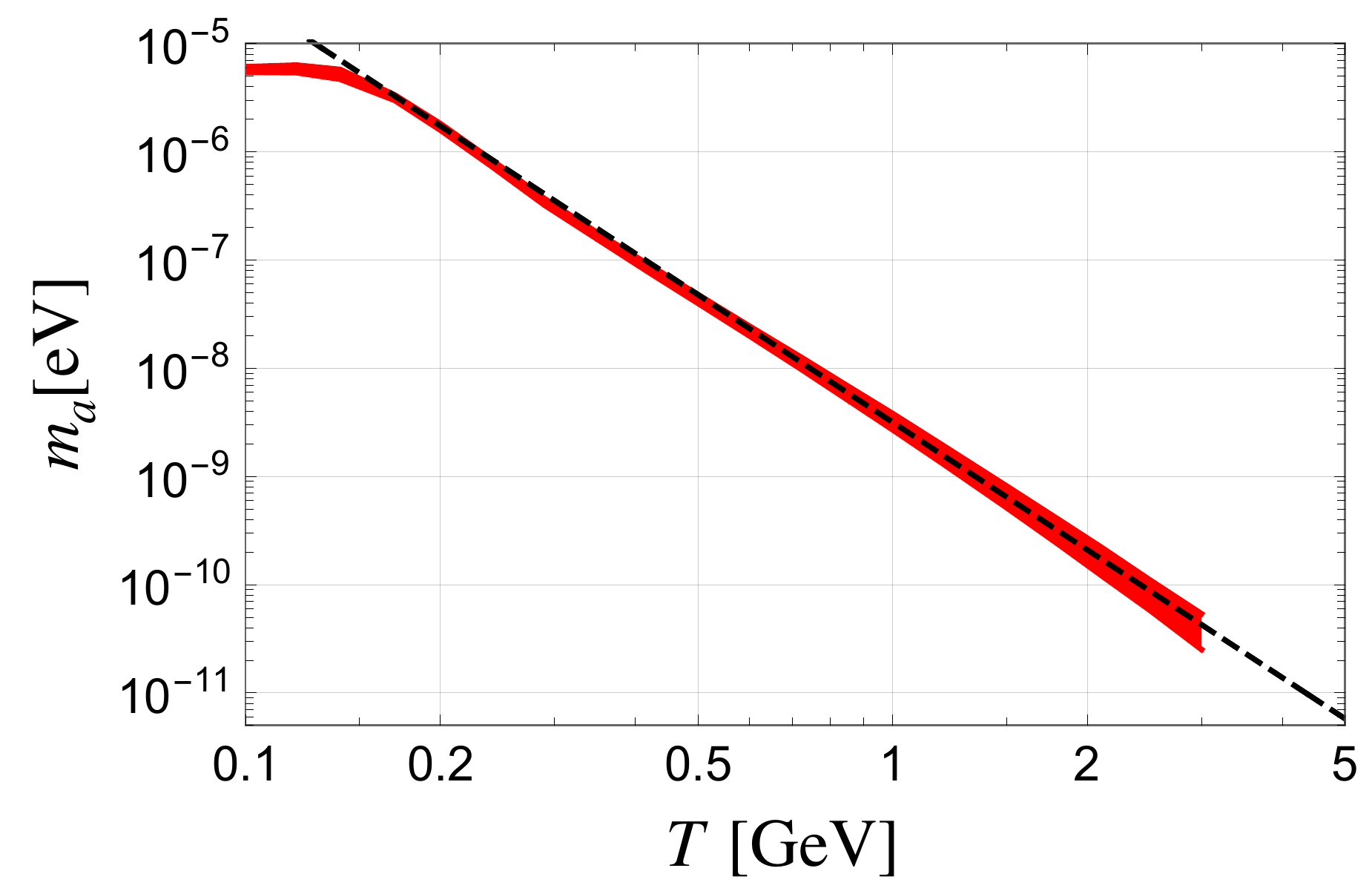}
\centering
\caption{
The temperature dependence of the axion mass for $f_a=10^{12}\GeV$ from the lattice result (red line)~\cite{Borsanyi:2016ksw}.
The width of the red line represents the statistical and systematic errors. 
The black dashed line represents the result of fitting from $1\GeV$ to $3\GeV$ with the power-law function (\ref{fitmass}).
}
\label{latticemass}
\end{figure}

At  temperatures comparable to or higher than $T_{\rm{QCD}}$,  the chiral perturbation theory is no longer viable, and the system should be described by the quark-gluon plasma rather than hadrons.
In this regime, we should rely on  non-perturbative methods like the lattice simulation. In Fig.~\ref{latticemass} we quote the latest lattice result~\cite{Borsanyi:2016ksw}
\footnote{The effective degrees of freedom $g_*$ and $g_{*s}$ \cite{Saikawa:2018rcs} we use in this work are also estimated by using this lattice result, $2+1+1$ QCD computation.} which covers the temperature region of $100\MeV<T<3\GeV$. The lattice result does not reach temperatures higher than $3$\,GeV, but the result from $1\GeV$ to $3\GeV$ is consistent with the dilute instanton gas approximation ($m^2_a(T)\propto T^{-b}$, $b=8.16$) which is applicable at such high temperatures. 
For later use, we fit the lattice result from $1\GeV$ to $3\GeV$ to obtain the temperature dependence of the axion mass,
\beq
m^2_a(T)\simeq1.0\times10^{-17}\eV^2\left(\frac{T}{1\GeV}\right)^{-n}\left(\frac{f_a}{10^{12}\GeV}\right)^{-2}\,
\label{fitmass}
\eeq
with $n=7.84$.
The fitting function (\ref{fitmass}) is shown by the black dashed line in Fig.~\ref{latticemass}.
We use this function to study the axion dynamics at temperatures higher than $T_{\rm QCD}$. Note that the axion mass around $1$\,GeV is most relevant for our purpose, and that the precise axion mass above $3$\,GeV does not alter our results as long as it decreases smoothly with increasing temperature.

Lastly, we comment on the potential shape. While the dilute instanton gas approximation at $T\gg T_{\rm{QCD}}$ gives the cosine-type potential, the potential obtained in the chiral perturbation theory at $T\ll T_{\rm{QCD}}$ is given by  (\ref{chiralpot}). For the parameter region of our interest, the temperature at the onset of oscillations is higher than $T_{\rm QCD}$, and therefore, we adopt the cosine potential with the temperature-dependent mass (\ref{fitmass}) in our numerical calculations. In fact, we have confirmed that the numerical results are almost the same for both potentials.

\subsubsection{Witten effect in the presence of monopoles
\label{sec:Witten}}
Let us first explain the Witten effect \cite{Witten:1979ey} in a hidden U(1)$_H$ gauge theory with a monopole. 
The relevant terms in the Lagrangian are
\beq
\mathcal{L}=-\frac{1}{4}F_{\mu\nu}F^{\mu\nu}-\frac{e_H^2\Theta}{64\pi^2}\epsilon_{\mu\nu\sigma\rho}F^{\mu\nu}F^{\sigma\rho}\,,
\eeq
where $e_H$ is the hidden gauge coupling constant, $F^{\mu\nu}$ is the field strength of the hidden gauge boson, and $\tilde{F}^{\mu\nu}$ is its dual.
The second term is physical and should not be omitted in the presence of a hidden monopole. The equation of motion is written as 
\beq
\del^\mu\left[F_{\mu\nu}+\frac{e_H^2\Theta}{16\pi^2}\epsilon_{\mu\nu\sigma\rho}F^{\sigma\rho}\right]=0\,.
\eeq
The second term modifies the Maxwell equation for the hidden gauge theory, which adds a new term to the Gauss's law,
\beq
\vec{\nabla}\cdot\vec{E}+\frac{e_H^2}{8\pi^2}\vec{\nabla}\cdot(\Theta\vec{B})=0\,,
\label{Gauss}
\eeq
where $E_i=F_{0i}$ and $B_i=-1/2\epsilon_{ijk}F^{jk}$ denote the hidden electric and magnetic fields, respectively.
When we introduce the hidden magnetic monopole with a magnetic charge $g_H$ ($=4\pi/e_H$), the Gauss's law for the magnetic field is given by
\beq
\vec{\nabla}\cdot\vec{B}=g_H(n_{M+}-n_{M-})\,,
\eeq
where $n_{M+(-)}$ is the number density of (anti-)monopoles.
From these two equations, we can understand that the monopole becomes a dyon, i.e. it acquires an electric charge of $e_H^2 g_H \Theta / (8 \pi^2) = e_H \Theta / (2\pi)$. This is called the Witten effect.

Now we shall introduce an axion coupled to the hidden gauge field.
Instead of the parameter $\Theta$, the axion is coupled to the gauge field as
\beq
\mathcal{L}_\theta=-\frac{e_H^2}{64\pi^2}\frac{(a- a_*)}{f_{a, H}}\epsilon_{\mu\nu\sigma\rho}F^{\mu\nu}F^{\sigma\rho}\,,
\eeq
where $f_{a, H}$ is the decay constant of the axion associated with U$(1)_H$ and $a_*$ is a constant. The decay constant $f_{a, H}$ is  of the same order of $f_a$ in an ordinary set-up, and they are related by $f_{a, H}=(N_{\rm{DW}}/N_{H})f_a$, where the domain wall numbers $N_{\rm{DW}}$ and $N_{H}$ are determined by the color and hidden U(1) anomaly coefficients, respecitvely. 
The axion-dependent contribution to the electromagnetic energy of a single monopole is estimated to be~\cite{Fischler:1983sc}
\beq
V_M&\sim&\beta f_{a, H}\frac{(a-a_*)^2}{f_{a, H}^2},\\
\beta&=&\frac{\alpha_H}{32\pi^2}\frac{1}{r_cf_{a, H}}\,,
\eeq
where $\alpha_H\equiv e_H^2/4\pi$, and $r_c$ is the radius of the monopole core.
When we consider a 't Hooft-Polyakov monopole~\cite{tHooft:1974kcl, Polyakov:1974ek}, $r_c$ is the inverse of the heavy gauge boson mass, $m_W$, which is about $\alpha_H$ times the monopole mass. 
Taking a spatial average over the whole space, 
we obtain the energy density of the axion ground state in the plasma with monopoles and anti-monopoles as $U=n_MV_M$, where $n_M=n_{M+}+n_{M-}$.
Thus, the homogeneous mode of the axion effectively obtains a mass of
\beq
m^2_{a, M}(T)
&=&2\beta\frac{n_M(T)}{f_{a, H}}
\\
&=&\frac{\alpha_H^2}{16 \pi^2}\frac{\rho_M(T)}{f_{a, H}^2}\,.
\label{Wittenmass}
\eeq
The mass (\ref{Wittenmass}) decreases in time due to the cosmic expansion, so that this does not spoil the Peccei-Quinn mechanism to solve the strong CP problem.

Finally, we comment on an upper bound on the coupling constant $\alpha_H$. 
The SU(2)$_H$ instanton effect gives another effective mass to the axion, which is proportional to $e^{-\pi /\alpha_H}$~\cite{Fuentes-Martin:2019bue,Csaki:2019vte,Buen-Abad:2019uoc} and does not decrease in time at a low temperature. 
In order to solve the strong CP problem by the PQ mechanism, the effective mass from this contribution should be smaller by a factor of $10^{-10}$ than the effective mass from the QCD effect. 
This requires 
\beq
\alpha_H\lesssim0.07 \, . 
\label{alphaH}
\eeq
We assume that this condition is satisfied throughout this paper.

\subsection{Dynamics of axion 
\label{sec:Wittensuppress}}
Here we describe the dynamics of the axion, and  evaluate  important physical quantities such as the temperature at the onset of oscillations and the axion abundance for qualitative understanding.

We consider the axion potential of 
\beq
V(a)=m^2_a(T)f_a^2\left(1-\cos\frac{a}{f_a}\right)+\frac{1}{2}m_{a, M}^2(T)(a-a_*)^2\,.
\label{potential}
\eeq  
The mass term originating from the Witten effect has the minimum at $a = a_*$, which is generically deviated from the low-energy minimum at $a=0$ where the strong CP phase vanishes. 
We focus on a pre-inflationary scenario where the PQ symmetry is spontaneously broken during inflation, and assume that hidden monopoles are generated sometime after inflation when the phase transition takes place in the hidden sector. We also assume that, at $T\gg T_{\rm{QCD}}$, the second term dominates over the first one, and the temporal minimum of the potential is located at $a\simeq a_*$.

First, we define $\omega$ that is useful to parametrize the strength of the Witten effect: 
\beq
 \omega&\equiv&\alpha_H^2 \lmk \frac{N_{H}}{N_{\rm{DW}}} \rmk^2\Omega_Mh^2, 
 \\
 &\simeq& 5.9 \times 10^{-4} \lmk \frac{\alpha_H}{0.07} \rmk^2 
 \lmk \frac{N_{H}}{N_{\rm{DW}}} \rmk^2 \lmk \frac{\Omega_Mh^2}{0.12} \rmk\,, 
\eeq
where $\Omega_M\equiv\rho_M/\rho_{\rm{crit}}$ represents the monopole density parameter, and $h$ is the reduced Hubble constant. 
In addition, the sum of the axion and the monopole abundance should not be larger than the dark matter abundance, 
i.e.,
\beq
\Omega_ah^2+\Omega_Mh^2\lesssim\Omega_{\rm{obs}}h^2\simeq0.12\,, 
\label{obs}
\eeq
since the monopole is stable and behave like matter. 
Thus $\omega$ is smaller than $5.9 \times 10^{-4}$ unless $N_{H} > N_{\rm DW}$.

The axion starts to oscillate around the temporal minimum of the potential, $a=a_*$, when the Hubble parameter becomes comparable to the mass due to the Witten effect, $H(T_{\rm{osc}}) \simeq m_{a,M}(T_{\rm{osc}})$.
Here $T_{\rm{osc}}$ is the temperature at the onset of oscillations, and it is given by~\cite{Kawasaki:2015lpf}
\beq
T_{\rm{osc}}\simeq64\GeV
\lmk \frac{\omega}{0.12} \rmk \left(\frac{f_a}{10^{12}\GeV}\right)^{-2} \, .
\label{Tosc}
\eeq
The axion abundance induced by this process is given by 
\beq
\frac{n_a}{s}\simeq\frac{H_{\rm{osc}} \lmk \theta_{\rm{ini}} - \theta_* \rmk^2 f_{a}^2/2}{s(T_{\rm{osc}})}\,,
\eeq
where $H_{\rm{osc}} \equiv H(T_{\rm{osc}})$, $\theta_{\rm ini} \equiv a_{\rm ini} / f_a$ and $\theta_* \equiv a_* / f_a$ with $a_{\rm ini}$ being the initial axion field value. 
We assume $0\leq \theta_* < \theta_{\rm max} \equiv \pi$ without loss of generality, and $\theta_{\rm ini}$ and $\theta_*$ satisfy  $\abs{\theta_{\rm ini} - \theta_* } \le \pi (N_{\rm DW}/N_{H})$.
Thus, the axion density parameter is estimated by
\beq
\Omega^{(1)}_ah^2\simeq3\times10^{-4}
\lmk \theta_{\rm{ini}} - \theta_* \rmk^2
\lmk \frac{\omega}{0.12} \rmk^{-1} 
\left(\frac{f_a}{10^{12}\GeV}\right)^3\,.
\label{firstcontribution}
\eeq
Here the upper index $(1)$ of $\Omega_a$ implies that it represents the axion abundance induced by the Witten effect. This is to distinguish it from the contribution due to the violation of adiabaticity when the first term in the potential (\ref{potential}) becomes relevant.

As the temperature goes down, the potential from the Witten effect becomes comparable to that from non-perturbative QCD effects, and the temporal minimum starts to move toward $a= 0$. 
Solving $m^2_{a, M}(T_{\rm{shift}})\simeq m_{a}^2(T_{\rm{shift}})$, we obtain the temperature at this time, 
\beq
T_{\rm{shift}}\simeq0.8\GeV\left(\frac{g_{*s}(T)}{80}\right)^{-1/(3+n)}
 \lmk \frac{\omega}{0.12} \rmk^{-1/(3+n)}\,, 
 \label{T_shift}
\eeq
where we assume $T_{\rm{osc}}>T_{\rm{shift}}$, i.e.,
\beq
\lmk \frac{\omega}{0.12} \rmk^{q}
\left(\frac{f_a}{10^{12}\GeV}\right)^{-2}\left(\frac{g_*(T)}{80}\right)^{1/(3+n)}\gtrsim0.01\,
\label{oshshift}
\eeq
with $q \equiv (n+4)(n+3)$.
This gives the lower bound on the parameter $\omega$ for a given $f_a$.

If the axion can adiabatically follow the movement of the temporal minimum, then the production of extra axion oscillations are exponentially suppressed in this process. In Refs.~\cite{Kawasaki:2015lpf, Kawasaki:2017xwt, Nomura:2015xil},  the adiabatic suppression mechanism was assumed to work perfectly and such a contribution was neglected. However, as we have seen in the previous section, it depends on the shape of the potential and the initial conditions whether the adiabatic suppression really works. In the following, we will study under what condition the adiabatic suppression works for the QCD axion.

\subsection{Numerical calculations of axion abundance
\label{sec:4-3}}
Here we solve the equation of motion for the axion taking account of the Witten effect, to see if the adaibatic suppression works and to find a parameter space in which we can explain the observed dark matter density.

Substituting (\ref{potential}) to (\ref{generaleom}), we obtain the following equation of motion,
\beq
F^2(\tau)\frac{d^2\theta}{d\tau^2}+F(\tau)\left\{\frac{dF(\tau)}{d\tau}+\sqrt{g_*(\tau)}\tau^{-2}\right\}\frac{d\theta}{d\tau}+\frac{10\Mpl^2}{\pi^2T_n^4}m^2_a(\tau)\sin\theta\nonumber\\
+k \omega g_{*s}(\tau)\tau^{-3}(\theta-\theta_*)=0\,,
\eeq
where $\theta\equiv a/f_a$ and 
\beq
 k\equiv\frac{\rho_{\rm{crit}}h^{-2}T_n^3}{36\pi^2m_{a,0}^2f_a^2s_0}\,,
\eeq
where $s_0$ is the present entropy density. The time variable $\tau\equiv T_n / T$ is defined as before, where the normalization factor is now chosen as 
\beq
 T_n = \sqrt{ m_{a,0} \Mpl} \, .
\eeq
The axion density parameter $\Omega_a = \rho_a / \rho_{\rm crit}$ is obtained by evaluating the asymptotic value of the number to entropy ratio $n_a/s$.

\begin{figure}[t!]
\begin{minipage}[t]{8cm}
\includegraphics[width=8cm]{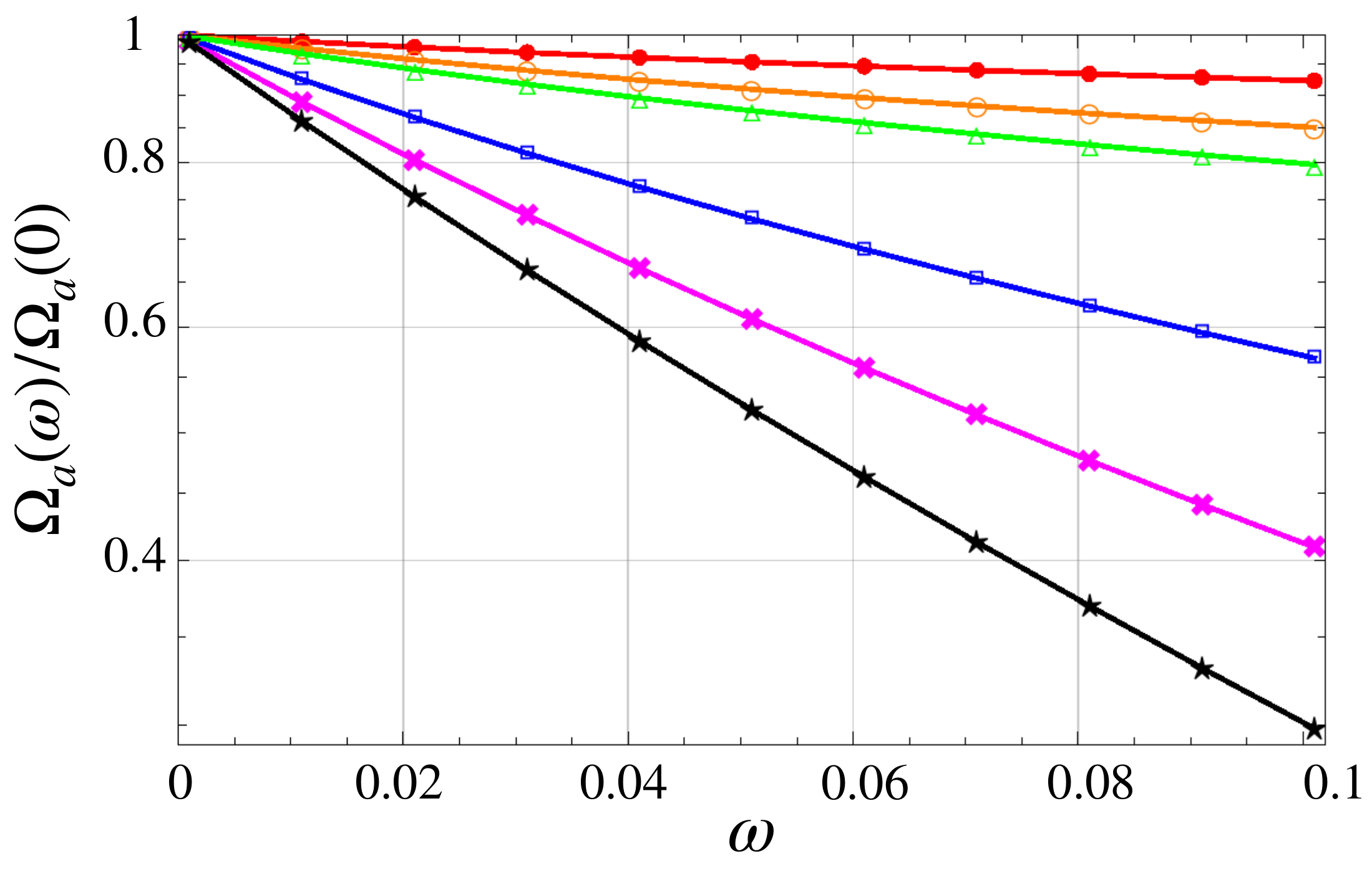}
\centering
\end{minipage}
\begin{minipage}[t]{8cm}
\includegraphics[width=8cm]{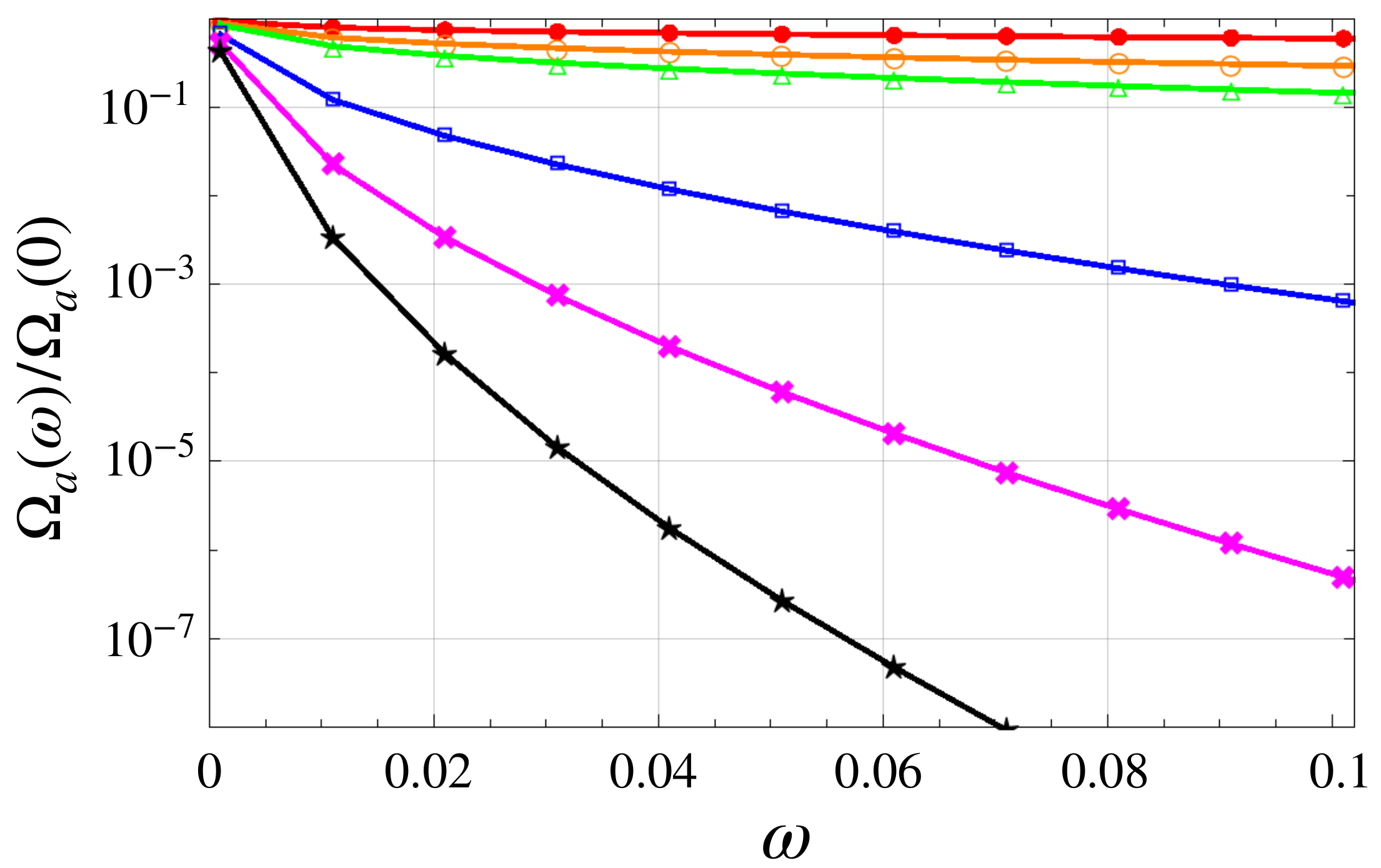}
\centering
\end{minipage}
\caption{The axion abundance normalized by the case without the Witten effect as a function of $\omega$ for various initial conditions, $\abs{\theta_* - \theta_{\rm max}} 
=0.01 \,({\rm red}\,\, \bullet),\, $
$0.05 \,({\rm orange}\,\, \circ),\,$
$0.1 \,({\rm green}\,\, \triangle),\,$
$0.5 \,({\rm blue}\,\, \square),\,$
$1\,({\rm magenta}\,\, \times)$, and
$\pi/2\,({\rm black}\,\, \star)$ from top to bottom.
We take $f_a=10^{12}$ (left panel) and $10^{11}\GeV$ (right panel).
}
\label{dependence}
\end{figure}

In order to see if the adiabatic suppression mechanism works, we plot the ratio between the density parameter with a given nonzero $\Omega_Mh^2$ and the one without it (i.e., with $\Omega_Mh^2=0$) in Fig.~\ref{dependence}. 
We take $\theta_{\rm{ini}} = \theta_*$ and $\del_\tau\theta_{\rm{ini}} = 0$ for the initial condition. 
In this case, the contribution from the first oscillation, \eq{firstcontribution}, is negligible because the initial field value is close to the temporal minimum of the potential.

Fig.~\ref{dependence} shows our results as a function of $\omega$ for $f_a=10^{12}$ and $10^{11}\GeV$, where the lines with $\bullet, \circ, \triangle, \square, \times, \star$ correspond to $\abs{\theta_* - \theta_{\rm max}} = 0.01, 0.05, 0.1, 0.5, 1, \pi/2$, respectively. 
As one can see from the left figure, the abundance is not strongly suppressed for $f_a = 10^{12} \GeV$. 
This is because the potential from the Witten effect is  weaker for a larger $f_a$ and the trapping effect is smaller.
In both figures, the abundance is  more suppressed for a larger $\omega$. 
This is because the axion is trapped by the Witten effect more strongly for a larger $\omega$ and the adiabatic suppression mechanism works more efficiently.

\begin{figure}[t!]
\begin{minipage}[t]{8cm}
\includegraphics[width=8cm]{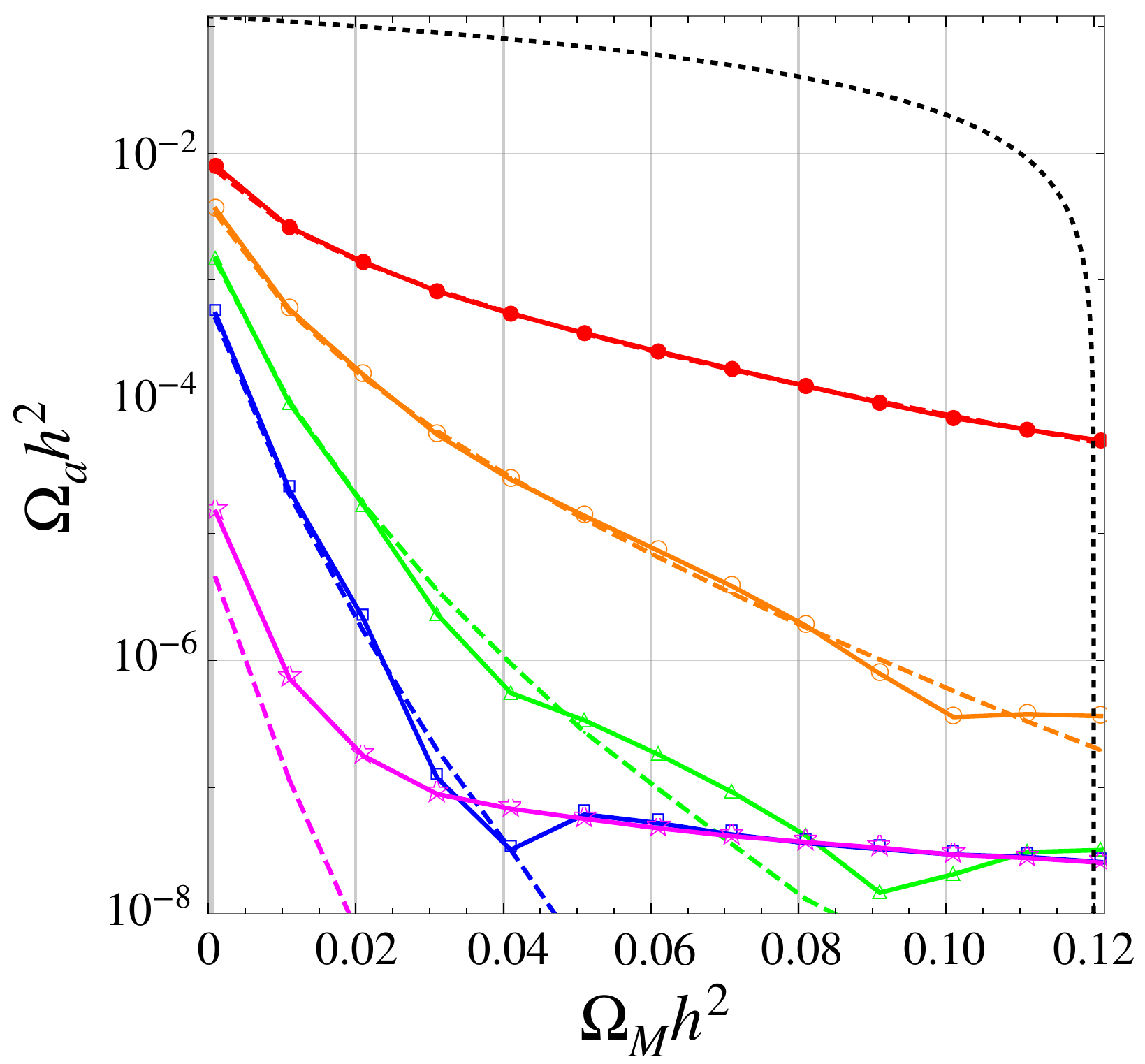}
\centering
\end{minipage}
\begin{minipage}[t]{8cm}
\includegraphics[width=8cm]{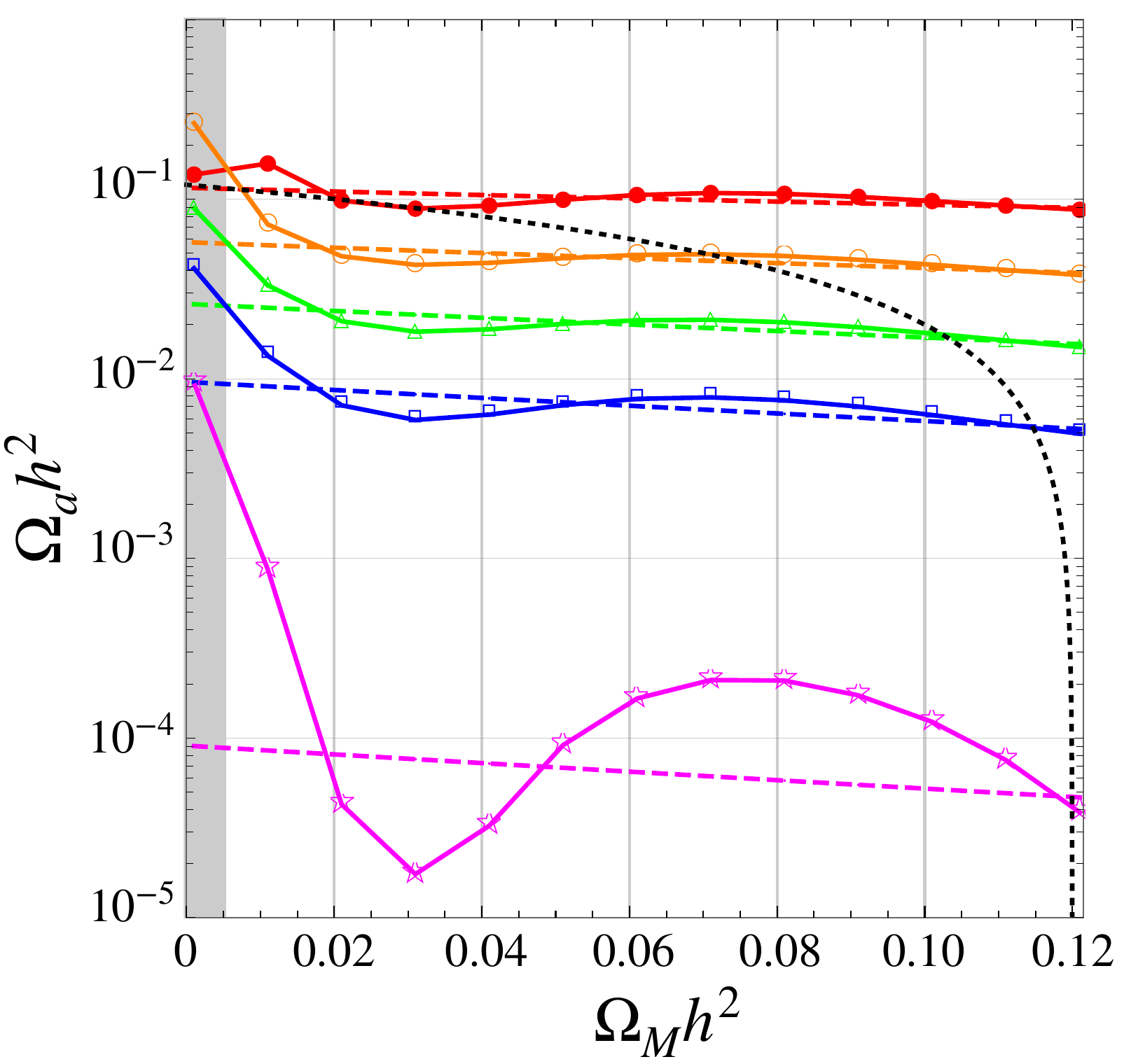}
\centering
\end{minipage}
\caption{
The axion abundance $\Omega_ah^2$ as a function of $\Omega_Mh^2$ 
for
$\abs{\theta_* - \theta_{\rm max}} =
0.5~({\rm red}~\bullet)$,\,
$1~({\rm orange}~\circ)$,\,
$\pi/2~({\rm green}~\triangle)$,\,
$\pi - 1~({\rm blue}~\square)$,
and $\pi - 0.1~({\rm magenta}~\star)$
from top to bottom.
We set $\alpha_H=0.07$ and  $f_a=10^{10}\GeV$ (left panel) and for $f_a=10^{11}\GeV$ (right panel).
Each solid line represents the solution for the initial condition with $\theta_{\rm{ini}}=\theta_*+(N_{\rm{DW}}/N_H)\, (=\theta_*+1)$, and each dashed line represents the one with $\theta_{\rm{ini}}=\theta_*$.
The black dotted line denotes $\Omega_a h^2 + \Omega_M h^2 = 0.12$.
We are not interested in the gray shaded region where $T_{\rm{osc}}\lesssim T_{\rm{shift}}$. 
}
\label{abundanceDW1}
\end{figure}

Next, we evaluate the density parameter for the axion $\Omega_ah^2$ to determine a realistic parameter space where both the axion and the monopole can explain dark matter.
The results for $N_{H}=N_{\rm{DW}}$ and $\alpha_H = 0.07$ are shown in Fig.~\ref{abundanceDW1} for $f_a=10^{10}\GeV$ (left panel) and $f_a=10^{11}\GeV$ (right panel), where the lines with $\bullet, \circ, \triangle, \square, \star$ correspond to the cases with $\abs{\theta_* - \theta_{\rm max}}=0.5, 1, \pi/2, \pi - 1, \pi - 0.1$, respectively.
Each solid line represents the solution for $\theta_{\rm{ini}}=\theta_*+ (N_{\rm DW}/N_H)$ ($= \theta_* + 1$), and each dashed line represents the one for $\theta_{\rm{ini}}=\theta_*$. 
In the former case, an oscillation is induced at $T = T_{\rm osc}$ and \eq{firstcontribution} may not be negligible. 
In the latter case, the contribution from the first oscillation, \eq{firstcontribution}, is negligible because the initial field value is close to the temporal minimum of the potential. 
The black dotted line represents the upper bound Eq.~(\ref{obs}).

The left panel of Fig.~\ref{abundanceDW1} shows that the axion abundance is strongly suppressed for $f_a=10^{10}\GeV$ and the axion is only a  subdominant component of dark matter.
For $\Omega_a h^2 \sim 10^{-(8\,\text{-}\,6)}$, (i.e., for small $\theta_*$ and large $\Omega_Mh^2$), the solid lines are deviated from the dashed lines. 
This is because the contribution from the first oscillation becomes important for the case of $\theta_{\rm{ini}}=\theta_*+1$ in that parameter space. 
Substituting $\theta_{\rm{ini}}=\theta_*+1$ into Eq.~(\ref{firstcontribution}), we obtain $\Omega_a^{(1)}h^2 \simeq 6 \times 10^{-8} (\Omega_M h^2/0.12)^{-1}$ for $f_a = 10^{10} \GeV$. 
In particular, this is independent of $\abs{\theta_* - \theta_{\rm max}}$. 
These can be actually observed in the left panel of Fig.~\ref{abundanceDW1}, where the magenta and blue lines are overlapped and satisfy $\Omega_a^{(1)}h^2 \sim 6 \times 10^{-8} (\Omega_M h^2/0.12)^{-1}$ for $\Omega_M h^2 \gtrsim 0.05$. 
One can also see that the green line become overlapped for $\Omega_M h^2 \gtrsim 0.1$. 
This means that the first oscillation around the potential by the Witten effect is dominant in this parameter space, while the second oscillation around the QCD potential is dominant in the other parameter space. 

In the right panel, the solid lines are deviated from the dashed lines at a small $\Omega_M h^2$. 
This is because the adiabatic suppression mechanism is not efficient as $f_a$ is larger than in the left panel. For a small $\Omega_M h^2$, the condition \eq{oshshift} is barely satisfied and the dynamics of the axion depends non-trivially on the initial condition. 
In the gray-shaded region, the condition \eq{oshshift} is violated and we cannot distinguish between the timing of oscillation by the Witten effect and the QCD effect.

\begin{figure}[t!]
\includegraphics[width=8cm]{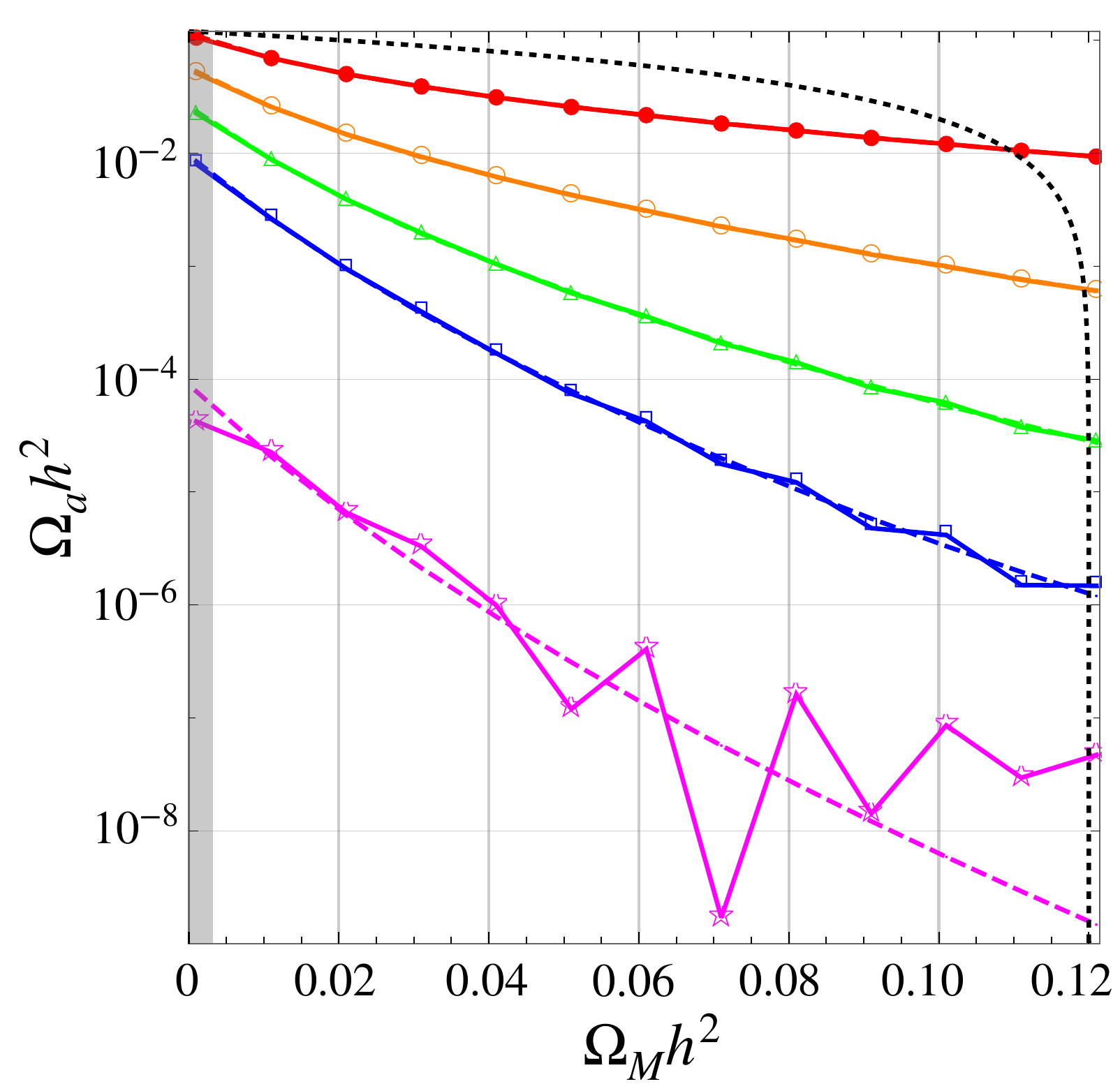}
\centering
\caption{
Same as the right panel of Fig.~\ref{abundanceDW1} 
(i.e., $f_a = 10^{11} \GeV$, $\alpha_H = 0.07$, and $\abs{\theta_* - \theta_{\rm max}}=0.5, 1, \pi/2, \pi - 1, \pi - 0.1$) but with $N_{H}/N_{\rm{DW}}=5$.
}
\label{abundanceDW5}
\end{figure}

Comparing the two panels in Fig.~\ref{abundanceDW1}, we can see that the axion abundance becomes larger as $f_a$ increases because the Witten effect is more suppressed. In the right panel, the axion abundance can be comparable to or even larger than the monopole abundance.
On the other hand, the axion of $f_a\gtrsim10^{11}\GeV$ can be adiabatically suppressed if we take $N_{H}>N_{\rm{DW}}$. 
This can be seen from Fig.~\ref{abundanceDW5}, which shows the axion abundance in the case of $N_{H}/N_{\rm{DW}}=5$ and $f_a=10^{11}\GeV$. 
The initial condition is taken to be $\theta_{\rm{ini}}=\theta_*$ for each solid line and $\theta_{\rm{ini}}=\theta_*+ (N_{\rm DW}/N_H)$ for each dashed line.

The reason why the solid lines are deviated from the dashed lines in Fig.~\ref{abundanceDW5} is the same as the one in the left panel in Fig.~\ref{abundanceDW1}. 
Namely, the contribution \eq{firstcontribution} becomes important for a very small $\Omega_a h^2$. 
However, one can see an oscillating behavior in Fig.~\ref{abundanceDW5}. 
The deviation appears when the temporal minimum starts to shift around the time when the axion oscillates due to the Witten effect, namely for the case of $T_{\rm{osc}} \sim T_{\rm{shift}}$. 
In this case, the oscillation amplitude at $T=T_{\rm{shift}}$ has a non-negligible effect on the axion dynamics when the temporal minimum is moving.
If the axion moves toward the same (opposite) direction as the temporal minimum, the abundance gets suppressed (enhanced). 
The result hence depends on the axion field value at $T=T_{\rm{shift}}$, which non-trivially depends on $\Omega_M h^2$.

\begin{figure}[t!]
\includegraphics[width=10 cm]{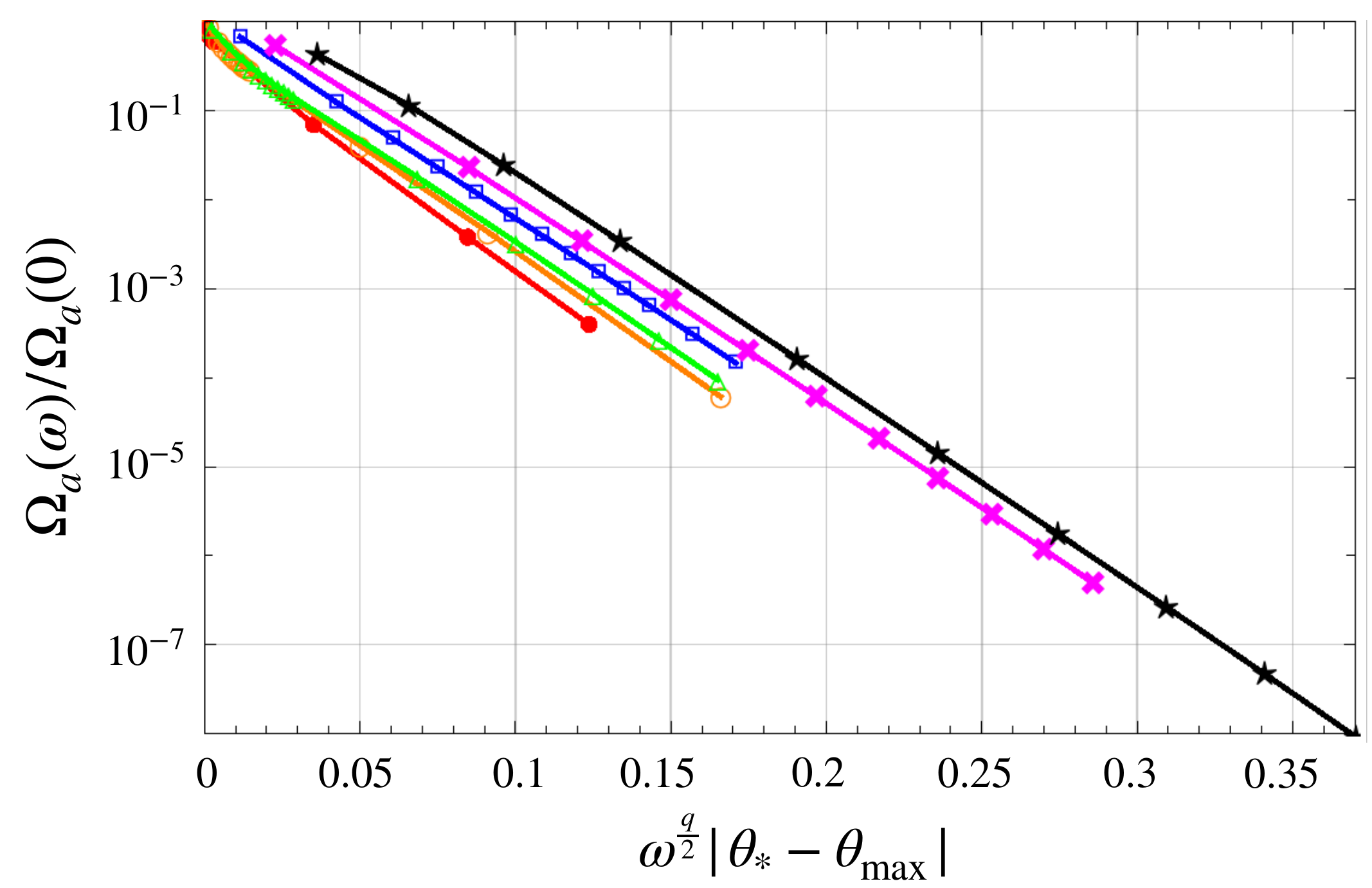}
\centering
\caption{
Same as the right panel of Fig.~\ref{dependence} (i.e., $f_a = 10^{11} \GeV$ and $\abs{\theta_* - \theta_{\rm max}}= 0.01, 0.05, 0.1, 0.5, 1, \pi/2$) but the horizontal axis is given by
the inverse of the parameter for the violation of adiabaticity,
$\omega^{q/2} \abs{\theta_* - \theta_{\rm max}}$.
}
\label{abundanceDW11log}
\end{figure}

To summarize the above argument, the adiabatic suppression mechanism works for $f_a \lesssim 10^{11} \GeV$ and large $\omega, \abs{\theta_* - \theta_{\rm max}}$.
For a qualitative understanding of the suppression factor in this case, we plot the ratio of the density parameter with and without the Witten effect as a function of $\omega^{q/2} \abs{\theta_* - \theta_{\rm max}}$ in Fig.~\ref{abundanceDW11log}. 
We can fit the result by an exponential function such as 
\beq
 \log \lkk \frac{\Omega_a (\omega) }{\Omega_{a} (0)} \rkk \propto 
 - \omega^{q/2} \abs{\theta_* - \theta_{\rm max}}\,.
 \label{resultaxion}
\eeq
Note that, while the lines in the figure are not completely overlapped, all of them exhibit the same exponential dependence like \eq{resultaxion}.
Their differences are only ${\cal O}(1)$ factors and hence the exponential dependence of \eq{resultaxion} is robust. In the next subsection, we will explain that this exponential suppression factor is related to the parameter for the violation of adiabaticity.

\subsection{Analytical evaluation of violation of adiabaticity
\label{sec:abundance}}
In this section, we give an analytic derivation of the suppression factor in \eq{resultaxion} using a similar analysis used in Sec.~\ref{sec:3}. 

We calculate the parameter for the violation of adiabaticity $\epsilon$ defined by \eq{adiabaticity} with $m^2_{\rm{eff}}\equiv |V''(a)|$.
Let us focus on the potential around the temporal minimum, which can be written as 
\beq
V(\phi) = \frac12 \lmk m_{a,M}^2 (T) - m_a^2(T) \rmk \lmk a - a_{\rm min}^{\rm temp} \rmk^2 
+ \frac{1}{24} \frac{m_a^2(T)}{f_a^2}  a^4
+ \dots\,, 
\eeq
where the dots represent higher-order terms in terms of $a/f_a$ and we define the temporal minimum as 
\beq
 a_{\rm min}^{\rm temp} 
 \equiv \frac{m_{a,M}^2 (T)}{m_{a,M}^2 (T) - m_a^2(T)} a_*\,.
 \label{delta2}
\eeq
The effective mass squared at a field value $a$ is given by 
\beq
 m_{\rm{eff}}^2(a) = m_{a,M}^2 (T) - m_a^2(T) + \frac{1}{2} \frac{m_a^2(T)}{f_a^2} a^2\,.
\eeq
As $T$ approaches $T_{\rm shift}$, the first and second term becomes cancelled and the third term becomes relevant. 
If we take $a = a_{\rm min}^{\rm temp}$, the threshold of the temperature $T_{\rm th}$ below which the third term becomes relevant is given by 
\beq
 m_{a,M}^2 (T_{\rm th}) - m_a^2(T_{\rm th}) &=& \lmk \frac{1}{2} \frac{m_a^2(T_{\rm th})}{f_a^2} m_{a,M}^4 (T_{\rm th}) a_*^2 \rmk^{1/3}\,. 
 \label{eq2}
\eeq

Now we shall calculate the parameter for the violation of adiabaticity $\epsilon$, assuming that it is sufficiently small. 
From this assumption, we expect that $a$ follows its temporal minimum, $a_{\rm min}^{\rm temp}$, until $T = T_{\rm th}$. 
Then it is calculated as 
\beq
\epsilon(T)
 &\simeq& \abs{ \frac{1}{\abs{m_{\rm eff}}^3} 
  \frac{1}{2t} 
\lmk -3 m_{a,M}^2 (T) - n m_{a}^2 (T) 
 \rmk}\,, 
\eeq
for $T \gtrsim T_{\rm th}$. This is maximum at $T \sim T_{\rm th}$ and 
\beq
 \epsilon \vert_{T = T_{\rm th}} 
 &\simeq& 
 \frac{\lmk 3 +n \rmk f_a}{\sqrt{2} t_{\rm th} m_a(T_{\rm th}) a_*}
 \\
  &\propto&
  T_{\rm shift}^{1/2} \omega^{-1/2} \abs{\theta_* - \theta_{\rm max}}^{-1}
  \\
  &\propto&
  \omega^{-q/2} \abs{\theta_* - \theta_{\rm max}}^{-1}\,,
\eeq
where we use \eq{eq2} and $a_* / f_a = \abs{\theta_* - \theta_{\rm max}}$. 
The last line comes from \eq{T_shift}. When $\epsilon \ll 1$, the resulting abundance is exponentially suppressed by a factor of $e^{-{\cal O}(1) / \epsilon}$, so that the power in the exponent is proportional to $\omega^{q/2} \abs{\theta_* - \theta_{\rm max}}$, which is actually consistent with our numerical result.

\section{Discussion and conclusions
\label{sec:conclusion}}
We have studied the scalar abundance in the presence of a trapping effect especially near the top of the low-energy potential. 
As the onset of oscillations is delayed by the trapping effect, one may naively expect that the scalar abundance is enhanced for a stronger trapping effect. This is indeed the case if the scalar field is trapped exactly at the top of the potential, like the flaton in the thermal inflation. However, this is not the case for a singlet scalar field without any special point in the potential, like the modulus field in the adiabatic suppression mechanism.

We have first investigated a toy model with a double-well potential and a Hubble-induced mass term. 
We have found that the scalar abundance can be logarithmically enhanced only if the scalar field is trapped very near the top of the potential and if the trapping effect is not too strong. 
For the case with a very strong trapping effect, the scalar abundance is exponentially suppressed because of the adiabatic suppression mechanism, even if the scalar field is trapped very near the top of the potential. 
The threshold is roughly given by $C \gtrsim \phi_{\rm min} / \phi$, where $C$ is the coefficient of the Hubble-induced mass and $\phi_{\rm min}$ is the field value at the minimum of the double-well potential. We have shown that this condition is related with the violation of the adiabaticity of parameters in the scalar potential, $|(dm_{\rm{eff}}/dt)/m^2_{\rm{eff}}|$.

We have also considered the QCD axion model with an effective potential due to the Witten effect in the presence of a monopole. 
We have clarified the condition for the adiabatic suppression mechanism to work in the vicinity of the top of the potential.\footnote{Such initial condition can be realized if the inflaton mixes with the axion, and shift the potential by an amount of $\pi$~\cite{Daido:2017wwb,Takahashi:2019pqf, Takahashi:2019qmh, Nakagawa:2020eeg}. (See also \cite{Co:2018mho, Kobayashi:2019eyg,Huang:2020etx}).}
The condition can be understood again by the violation of the adiabaticity of parameters in the scalar potential. 
Compared with the previous works~\cite{Kawasaki:2015lpf,Kawasaki:2017xwt}, we have added a condition to the gauge coupling $\alpha_H$ so that the SU(2)$_H$ instanton does not change the low-energy minimum of the QCD axion. 
Because of this condition, it is difficult to realize the adiabatic suppression for $f_a \gtrsim 10^{11}$\,GeV. However, the Witten effect can be stronger if the monopole abundance is larger than the observed DM abundance. This scenario is consistent with observations if the monopole can disappear before the BBN. Another possibility is to consider the case of $N_H \gg N_{\rm DW}$, which can be realized in the clockwork QCD axion model~\cite{Higaki:2015jag,Higaki:2016yqk}.

In summary, we have found that 
the scalar abundance is exponentially suppressed by the adiabatic suppression mechanism even if one strongly traps it very near the top of the low-energy potential. 
If the trapping effect is marginally strong, the scalar abundance can be logarithmically enhanced.
The only exception is that the top of the low-energy potential is a symmetry-enhancement point and one can trap the scalar field exactly at the top, like the flaton in the thermal inflation. 
If one considers an axion, which has a shift symmetry, or a singlet scalar field without any special point in its field space, the abundance is generically suppressed by the adiabatic suppression mechanism in the presence of a sufficiently strong trapping effect or at most logarithmically enhanced by choosing the initial condition extremely close to the top of the potential.

\section*{Acknowledgements}
S.N. acknowledges support from GP-PU at Tohoku University.
The present work is supported by JSPS KAKENHI Grant Numbers
17H02878 (F.T.), 20H01894 (F.T.),
20H05851 (F.T. and M.Y.), JP20K22344 (M.Y.), 
World Premier International Research Center Initiative (WPI Initiative), MEXT, Japan. 
M.Y. was supported by the Leading Initiative for Excellent Young Researchers, MEXT, Japan.

\bibliography{references}

\end{document}